\documentclass{article}
\usepackage[utf8]{inputenc}
\usepackage{bm,amssymb,amsmath,amsthm,natbib,float,color,enumitem}
\usepackage{fnpct}
\usepackage[colorlinks=true,citecolor=blue]{hyperref}

\theoremstyle{definition}
\newtheorem{theorem}{Theorem}
\newtheorem{claim}{Claim}
\newtheorem{proposition}{Proposition}

\usepackage[textwidth=520pt,textheight=600pt]{geometry}

\begin{document}
\title{Characterizations of Proportional Division Value in TU-Games \\ via Fixed-Population Consistency\thanks{We are grateful to William Thomson for his very helpful comments. We also thank the editor and the anonymous referees for their valuable comments, which helped us improve the paper. Nakada acknowledges financial support from the Japan Society for the Promotion of Science KAKENHI, Grant No.~25K16606. Koriyama and Tamura acknowledge financial support from Investissements d'Avenir, ANR-11-IDEX-0003/Labex Ecodec/ANR-11-LABX-0047.}}

\author{
Yukihiko Funaki\footnote{School of Political Science and Economics, Waseda University. E-mail:\texttt{funaki@waseda.jp}.}
\and
Yukio Koriyama\footnote{CREST, Ecole Polytechnique, Institut Polytechnique de Paris. E-mail:\texttt{yukio.koriyama@polytechnique.edu}.}
\and
Satoshi Nakada\footnote{School of Management, Department of Business Economics, Tokyo University of Science. E-mail:\texttt{snakada@rs.tus.ac.jp}.}
\and Yuki Tamura\footnote{CREST, Ecole Polytechnique, Institut Polytechnique de Paris. E-mail:\texttt{yuki.tamura@polytechnique.edu}.}
}
\date{\today}
\maketitle

\begin{abstract}
We study the proportional division value in TU-games, which distributes the grand coalition worth in proportion to the players' stand-alone worths. Focusing on \emph{fixed-population} consistency, we characterize the proportional division value through three types of axioms: axioms concerning transformations of the grand coalition worth, composition axioms, and nullified-game consistency. The first type captures how payoffs respond when the grand coalition worth changes. The composition axioms require that the outcome be the same whether a game is solved directly or through a two-step procedure based on a reference allocation and an associated adjustment of the game. Nullified-game consistency requires that, when some players' payoffs are fixed, the solution for the remaining players in the adjusted game coincides with their original payoffs. Together with efficiency and/or fairness-related conditions where appropriate, these axioms characterize the proportional division value.

\end{abstract}

\newpage
\section{Introduction}

The proportional division value is a simple and intuitive solution in cooperative games with transferable utility (TU-games): each player receives a share of the grand coalition worth in proportion to the player's stand-alone worth. Despite its intuitive appeal, it has received relatively less axiomatic attention compared to solutions such as the Shapley value \citep{Shapley1953} or other linear solutions.

In this paper, we focus on \emph{fixed-population} consistency properties satisfied by the proportional division value. We use the term fixed-population consistency to refer to relational requirements that compare the solution across games defined on the same set of players. This differs from the classical notion of variable-population consistency, where some players leave after receiving their payoffs and the player set therefore changes.\footnote{For a discussion of various interpretations of consistency, see \citet{thomson2012ep}.} In our approach, the underlying game may be modified in a structured way, while the set of players is unchanged. Such modifications may concern the grand coalition worth, the decomposition and recomposition of a game, or the construction of a residual game after some players' payoffs have been fixed. The common feature of these requirements is that the solution should respond systematically to such structured changes in the game. This perspective is natural in environments where the same group of agents remains involved while the total value generated by cooperation, or the way this value is represented, may change. From a technical standpoint, fixed-population consistency may require a distinct line of proof, since the techniques used in the variable-population case cannot be applied directly.

Our analysis proceeds along three lines. First, we study two axioms concerning transformations of the grand coalition worth. The grand coalition worth can be interpreted as the resource, surplus, or deficit to be allocated among the players. \emph{Grand-coalition homogeneity} requires that, when this worth is rescaled while all other coalition worths are kept fixed, each player's payoff is rescaled by the same factor. \emph{Grand-coalition normalization} requires that, when the sum of the players' stand-alone worths is subtracted from the grand coalition worth, each player's payoff decreases by exactly the player's stand-alone worth. These axioms compare games with the same worths for all coalitions other than the grand coalition and require the solution to respond systematically to changes in the aggregate value available to the grand coalition. Our first result states that these two axioms characterize the proportional division value (Theorem~\ref{thm:homogeneity}).

Second, we develop \emph{composition} axioms, extending the logic of composition axioms from bankruptcy problems to general TU-games. In bankruptcy problems, composition axioms describe how an allocation should respond when the amount of the resource to be divided is revised after an initial allocation has been determined. Translating this idea to TU-games, we interpret the grand coalition worth as the aggregate value to be allocated among the players. The composition axioms then compare two ways of determining the final payoff vector: applying the solution directly to the target game, or first applying it to a reference game and then applying it again to the residual game generated by the reference payoffs. Requiring these two procedures to lead to the same payoff vector captures a fixed-population form of consistency, since the player set is unchanged throughout the comparison. Our results show that this composition logic, suitably adapted to TU-games, characterizes the proportional division value on the general domain and on a bankruptcy-type domain (Theorems~\ref{thm:composition} and~\ref{thm:young}).

Finally, we study a \emph{nullified-game consistency} axiom. This axiom considers a situation in which the payoffs of some players are taken as fixed and asks how the solution should treat the remaining players. The original game is adjusted by neutralizing the role of the players whose payoffs have already been fixed, while the player set itself is unchanged. The requirement is that the payoffs assigned to the remaining players in this adjusted game coincide with their original payoffs. Thus, the axiom provides another fixed-population consistency test: after some payoff obligations have been accounted for, applying the solution to the corresponding residual game should reproduce the payoffs of the players who remain active in the allocation problem. Combined with efficiency and a fairness-related condition, this axiom provides two characterizations of the proportional division value (Theorems~\ref{thm:nullifiedconsistencyer} and~\ref{thm:nullifiedconsistencyadd}).

Taken together, our results show that the proportional division value is not only simple and intuitive, but also robust to several forms of fixed-population consistency. The three groups of axioms studied in the paper formalize different ways in which a TU-game may be modified: the grand coalition worth may change, a game may be decomposed and recomposed, or some players' payoffs may be fixed and incorporated into a residual game. Across these different comparisons, the proportional division value is the unique solution satisfying the corresponding consistency requirements together with efficiency and/or fairness-related conditions where appropriate.

A possible criticism of the proportional division value is that it does not exploit the full information contained in a TU-game, most notably intermediate coalition worths, and therefore cannot capture complementarity or substitutability encoded in the characteristic function. This observation is valid. However, it reflects a trade-off between informational richness and simplicity. The proportional division value is parsimonious and transparent: it depends only on stand-alone worths and the grand coalition worth. This parsimony is attractive in applications where coalition values beyond small groups are difficult to interpret, costly to elicit, or statistically noisy. It is also natural in environments that resemble bargaining problems, where individual outside options and the aggregate surplus are more salient than the values of all intermediate coalitions. In such environments, a solution that relies only on individual stand-alone worths and the aggregate value generated by the grand coalition may provide a useful benchmark.

The proportional division value and its variants have been studied in several settings, including universal TU-games and important subclasses of TU-games.\footnote{Examples include \cite{kk2015td}, \cite{ortmann2000mmor}, \cite{kd2003mmor}, \cite{zvcf2021scw}, \cite{zvf2022td}, and \cite{fk2025ijgt}.} For a comprehensive overview, we refer to \cite{zou2021thesis}. Complementing this literature, \cite{fk2025ijgt} characterize the proportional division value and related variants under the individual monotonicity axiom, allowing shares to deviate from stand-alone worths and, through zero-sum adjustments, to incorporate redistributive social objectives. Separately, \cite{fknt2025wp} study a different class of solutions, namely linear solutions, and provide characterizations based on composition axioms and nullified-game consistency axioms. Our contribution differs from these works by focusing on the proportional division value itself and by showing how it can be characterized through several fixed-population consistency requirements.

Our results are also connected to several contributions in the literature on bankruptcy problems. \citet{young1987mor} characterizes the broader class of parametric rules by equal treatment of equals, continuity, and consistency. Several subsequent papers study composition axioms, which describe how rules respond when the available resource changes. \citet{moulin2000econometrica} characterizes rules satisfying consistency, upper and lower composition,\footnote{These correspond, respectively, to composition down and composition up in our paper.} and scale invariance, and \citet{chambers2006geb} extends this analysis by characterizing difference rules through composition up and composition down. \citet{chambersmorenoternero2017scw} show that composition down, together with continuity, equal treatment of equals, and consistency, characterizes a broad family of generalized equal-sacrifice rules, which contains the equal-sacrifice rules of \citet{young1988jet} as a special case and also includes constrained equal-sacrifice and exogenous poverty-line rules. These results provide important benchmarks for our composition axioms, which adapt related ideas to the TU-game setting.

Another closely related line of work studies fixed-population counterparts of consistency. \citet{dtt2024scw} introduce \emph{partial implementation invariance}, a fixed-population counterpart of consistency, and study how characterizations based on variable-population consistency are preserved or modified when the population is kept fixed. This idea is close in spirit to our nullified-game consistency axiom in Section~\ref{sec:nullifiedconsistency}: in both cases, some payoffs are treated as already accounted for, the problem is adjusted accordingly, and the solution is required to reproduce the payoffs of the remaining agents. The TU-game setting, however, requires different formulations because coalition values other than the grand coalition worth are also part of the model. Sections~\ref{sec:composition} and~\ref{sec:nullifiedconsistency} make these connections explicit by adapting composition and fixed-population consistency ideas from bankruptcy problems to the richer environment of TU-games.

The rest of the paper is organized as follows. In Section~\ref{sec:model}, we define the model. In Section~\ref{sec:homogeneity}, we establish a characterization of the proportional division value based on grand-coalition homogeneity and grand-coalition normalization. In Section~\ref{sec:composition}, we provide characterizations based on composition axioms and discuss the relationship between our results and characterizations of bankruptcy rules, including the proportional solution. In Section~\ref{sec:nullifiedconsistency}, we present characterizations based on nullified-game consistency and relate this axiom to fixed-population consistency ideas in bankruptcy problems. Finally, in Section~\ref{sec:conclusion}, we conclude.

\section{The Model}\label{sec:model}
A cooperative game with transferable utility (TU-game) is a pair $(N,v)$, where $N$ is a finite set of players with $|N| \geq 2$ and $v: 2^N \to \mathbb{R}$ is a characteristic function satisfying $v(\emptyset) = 0$. For each $S \subseteq N$, $v(S)$ represents the \emph{worth} of coalition~$S$. To simplify notation, we write $v(i)$ for the stand-alone worth $v(\{i\})$ of player $i$.
Throughout the paper, we fix the player set $N$ and identify each game with its characteristic function. Let $\mathcal{V}^{all}$ denote the set of all TU-games, and let $\mathcal{V} \subseteq \mathcal{V}^{all}$ denote the subclass of games with $\sum_{k\in N} v({k}) \neq 0$. Our analysis focuses on $\mathcal{V}$, except in Section \ref{sec:nullifiedconsistency}, where we consider subgames of $\mathcal{V}$, and in Lemma \ref{lemma:gn&l} (Appendix \ref{appendix:incompatible}), which applies to $\mathcal{V}^{all}$.  For each $v \in \mathcal{V}$ and each $\alpha \in \mathbb{R}$, let $v^\alpha \in \mathcal{V}$ be defined by 
$$
v^\alpha(S) :=
\begin{cases}
v(S) \ &\text{if} \ S \neq N, \\
\alpha \ &\text{if} \ S = N. 
\end{cases}
$$

A solution is a function $\varphi: \mathcal{V} \to \mathbb{R}^N$, where for each $v \in \mathcal{V}$ and each $i \in N$, $\varphi_i(v)$ is the \emph{payoff} assigned to player $i$ in~$v$.\footnote{While the term \emph{solution} is sometimes used for set-valued mappings, throughout this paper we use it exclusively for single-valued mappings.} We focus on the \emph{proportional} division value: for each $v \in \mathcal{V}$ and each $i \in N$,
$$
P_i(v)=\frac{v(i)}{\sum_{k \in N}v(k)}v(N).
$$

\section{A Homogeneity Axiom}\label{sec:homogeneity}

We begin by establishing a characterization of the proportional division value based on a homogeneity axiom. This axiom concerns how a solution responds when the grand coalition worth is rescaled while all other coalition worths are kept fixed. It requires each player's payoff to be multiplied by the same factor as the grand coalition worth. Equivalently, the solution is homogeneous of degree one with respect to the grand coalition worth. Thus, the axiom captures scale invariance with respect to the aggregate value available to the grand coalition: no player's relative share changes merely because the overall ``size of the pie'' has been rescaled.

The axiom can also be viewed as a fixed-population consistency requirement. It compares two games with the same player set and the same coalition worths except for the grand coalition. Thus, no player is added or removed, and no smaller coalition is modified. The only change concerns the aggregate value assigned to the grand coalition, which can be interpreted as the resource, surplus, or deficit to be allocated. In this setting, consistency means compatibility between this structured change in the game and the induced change in payoffs: when the aggregate value is rescaled, the payoff vector is rescaled by the same factor. In this way, the axiom imposes a systematic response across a well-defined class of related games. Although some axioms studied in this paper can be formulated on larger domains, including $\mathcal{V}^{all}$, we state them on $\mathcal{V}$ when they are used in characterizations of the proportional division value, since $P$ is defined on $\mathcal{V}$.

\medskip
\noindent\textbf{Grand-Coalition Homogeneity}: For each $v \in \mathcal{V}$ and each $\alpha \in \mathbb{R}$,
\[
\varphi\left(v^{\alpha v(N)}\right) = \alpha \varphi(v).
\]

Several axioms related to \emph{grand-coalition homogeneity} have been studied in TU-games. The most fundamental is \emph{homogeneity}: if the worth of each coalition is multiplied by the same factor, then each player's payoff is multiplied by that factor as well. \citet{fk2025ijgt} define \emph{weak homogeneity}, which restricts this requirement to unanimity games. Together with other axioms, \emph{weak homogeneity} characterizes a family of solutions called \emph{weighted surplus sharing} values.

\citet{fk2025ijgt} also introduce \emph{weak grand-coalition homogeneity}, a weaker form of \emph{grand-coalition homogeneity} that applies only to unanimity games. Together with other axioms, this axiom characterizes the \emph{shifted proportional division value}, a variant of the proportional division value in which each player's stand-alone worth is modified by a zero-sum adjustment, and the resulting modified worth determines the player's share.

While homogeneity-type axioms concern robustness with respect to scaling coalitional worths, \emph{fairness}, introduced by \citet{van2002ijgt}, and \emph{weak fairness}, introduced by \citet{vcfp2016td}, concern additive changes: modifying the worth of a coalition changes the payoffs of all players, or of a specified group of players, by the same amount.

Our next axiom concerns a different transformation of the grand coalition worth. It states that subtracting the sum of the players' stand-alone worths from the grand coalition worth should reduce each player's payoff by the player's own stand-alone worth. The motivation is that the stand-alone worth $v(i)$ represents the value that player $i$ can generate independently, and therefore provides a natural baseline in allocating the grand coalition worth. The transformation in \emph{grand-coalition normalization} removes the aggregate stand-alone component $\sum_{k\in N}v(k)$ from the grand coalition worth, while leaving all other coalition worths unchanged. The axiom requires this aggregate adjustment to be reflected in payoffs through the corresponding individual baselines: player $i$'s payoff is reduced by $v(i)$. Thus, the axiom induces a baseline--surplus decomposition, separating the individualistic component $v(i)$ from the cooperative component $v(N)-\sum_{k\in N}v(k)$.

This interpretation is consistent with bargaining and contractual settings in which outside options are secured first and negotiations concern how to share the remaining gains or losses. It also highlights the informational simplicity of the proportional division value: the axiom treats stand-alone worths as baseline components and leaves only the cooperative surplus or deficit generated by the grand coalition to be allocated.

\medskip
\noindent\textbf{Grand-Coalition Normalization}: For each $v \in \mathcal{V}$ and each $i \in N$,
\[
\varphi_i\left(v^{v(N)-\sum_{k \in N}v(k)}\right) = \varphi_i(v) - v(i).
\]

\emph{Grand-coalition normalization} is related to \emph{covariance}, a standard invariance axiom for TU-game solutions. \emph{Covariance} requires that, if a game $v$ is transformed into $v'=\alpha v+\sum_{i\in N}\beta_i u_i$ with $\alpha>0$ and $\beta\in\mathbb{R}^N$, then the solution transforms accordingly, that is, $\varphi_i(v')=\alpha \varphi_i(v)+\beta_i$.\footnote{For each $T\subseteq N$, the game $u_T$ is called the $T$-unanimity game: $u_T(S)=1$ if $T\subseteq S$, and $0$ otherwise.} In particular, when $\alpha=1$ and $\beta_i=-v(i)$ for each $i\in N$, \emph{covariance} implies
$\varphi_i\!\left(v-\sum_{i\in N}v(i)u_i\right)=\varphi_i(v)-v(i)$.
Thus, \emph{covariance} prescribes how payoffs adjust when stand-alone components are subtracted from the whole game. By contrast, \emph{grand-coalition normalization} imposes an analogous payoff adjustment under a more limited transformation: only the grand coalition worth is changed, by subtracting the sum of the stand-alone worths.

This comparison illustrates how different invariance requirements lead to different, yet structurally simple, solution concepts. A simple solution that depends only on the grand coalition worth and the stand-alone worths, and that satisfies \emph{covariance}, is the Center of the Imputation Set (CIS) value.\footnote{The Center of the Imputation Set (CIS) value \citep{DriessenFunaki1991} is defined by $CIS_i(v)=v(i)+\frac{1}{|N|}\left(v(N)-\sum_{k\in N}v(k)\right)$ for each $v\in\mathcal{V}$ and each $i\in N$.} Thus, the particular transformation under consideration matters: \emph{covariance} leads naturally to the CIS value, whereas the grand-coalition transformations studied here lead to the proportional division value.

\emph{Grand-coalition homogeneity} and \emph{grand-coalition normalization} are naturally considered together. The first axiom describes how payoffs respond when the grand coalition worth is rescaled. The second describes how payoffs respond when the aggregate stand-alone component is removed from the grand coalition worth. Together, they impose two complementary compatibility requirements between transformations of the grand coalition worth and the corresponding changes in payoffs. As Theorem~\ref{thm:homogeneity} shows, these two requirements characterize the proportional division value.

\medskip
\begin{theorem}\label{thm:homogeneity}
A solution $\varphi: \mathcal{V} \to \mathbb{R}^N$ satisfies \emph{grand-coalition homogeneity} and \emph{grand-coalition normalization} if and only if $\varphi = P$.
\end{theorem}
\vspace{-0.5cm}

\begin{proof}
$(\Leftarrow)$ Let $v \in \mathcal{V}$, $i \in N$, and $\alpha \in \mathbb{R}$.
\[
P_i\left(v^{\alpha v(N)}\right)
=
\frac{v(i)}{\sum_{k \in N} v(k)} \alpha v(N)
=
\alpha P_i(v).
\]
Hence, the proportional division value satisfies \emph{grand-coalition homogeneity}. Also,
\begin{eqnarray*}
P_i\left(v^{v(N)-\sum_{k \in N}v(k)}\right)
&=&
\frac{v(i)}{\sum_{k \in N} v(k)}
\left(v(N)-\sum_{k \in N}v(k)\right) \\
&=&
\frac{v(i)}{\sum_{k \in N} v(k)} v(N)
-
\frac{\sum_{k \in N} v(k)}{\sum_{k \in N} v(k)}v(i) \\
&=&
P_i(v)-v(i).
\end{eqnarray*}
Hence, the proportional division value satisfies \emph{grand-coalition normalization}.

\bigskip

\noindent $(\Rightarrow)$ Let $\varphi$ be a solution satisfying \emph{grand-coalition homogeneity} and \emph{grand-coalition normalization}.
Let $v \in \mathcal{V}$ and $i \in N$. By \emph{grand-coalition homogeneity}, when $v(N)=0$, $\varphi_i(v)=0=P_i(v)$. Now suppose that $v(N)\neq 0$. By \emph{grand-coalition homogeneity},
\begin{eqnarray}\label{eq:homogeneity}
\varphi_i\left(v^{v(N)-\sum_{k \in N} v(k)}\right)
&=&
\varphi_i\left(v^{\frac{v(N)-\sum_{k \in N}v(k)}{v(N)}v(N)}\right) \nonumber\\
&=&
\left(\frac{v(N)-\sum_{k \in N} v(k)}{v(N)}\right)\varphi_i(v).
\end{eqnarray}
By \emph{grand-coalition normalization} and \eqref{eq:homogeneity},
\[
\varphi_i(v)
=
v(i)+\varphi_i\left(v^{v(N)-\sum_{k \in N}v(k)}\right)
=
v(i)+
\left(\frac{v(N)-\sum_{k \in N} v(k)}{v(N)}\right)\varphi_i(v).
\]
Then
\[
\varphi_i(v)
=
\frac{v(i)}{\sum_{k \in N} v(k)}v(N)
=
P_i(v).
\]
\end{proof}

Before turning to independence, we note the relation between these axioms and \emph{efficiency}.\footnote{The definition is provided in Section \ref{sec:nullifiedconsistency}.} Neither \emph{grand-coalition homogeneity} nor \emph{grand-coalition normalization} implies \emph{efficiency} by itself. For example, the zero solution on $\mathcal{V}$ satisfies \emph{grand-coalition homogeneity} but is not \emph{efficient} whenever $v(N)\neq 0$. For each fixed vector $c\in\mathbb{R}^N$, the solution $\varphi(v)=P(v)+c$ on $\mathcal{V}$ satisfies \emph{grand-coalition normalization}, but it is not \emph{efficient} whenever $\sum_{i\in N}c_i\neq 0$. Theorem~\ref{thm:homogeneity} shows that \emph{efficiency} is implied by the combination of the two axioms, since the combination characterizes the proportional division value.

We provide examples to demonstrate the independence of the axioms used in Theorem~\ref{thm:homogeneity}. First, consider a dictatorship, where one player always receives the entire the grand coalition worth, and all other players receive nothing. This solution satisfies \emph{grand-coalition homogeneity} but violates \emph{grand-coalition normalization}.

Next, we drop \emph{grand-coalition homogeneity}. Let $|N|=n$, and for notational convenience, label the players so that $N=\{1,\dots,n\}$. Let $\mathcal{W}=\{v \in \mathcal{V} : \text{for each } i \neq n,\ v(i)=0 \}$. Because $\mathcal{W}\subseteq\mathcal{V}$, each $v\in\mathcal{W}$ satisfies $v(n)=\sum_{k\in N}v(k)\neq 0$. Let $x:=(x_i)_{i=1}^{n-1}\in\mathbb{R}^{n-1}$ be such that $\sum_{i=1}^{n-1}x_i=0$. Define the solution $\varphi$ as follows: for each $v\in\mathcal{W}$, set $\varphi(v)=(x,v(N))$,\footnote{This means that each player $i \in \{1,\dots,n-1\}$ is assigned $x_i$, and player $n$ is assigned $v(N)$.} and for each $v\in\mathcal{V}\setminus\mathcal{W}$, set $\varphi(v)=P(v)$. The solution violates \emph{grand-coalition homogeneity}: for each $v\in\mathcal{W}$ and $i=1,\dots,n-1$, $\varphi_i(v)$ is fixed at $x_i$, so these players' payoffs are independent of $v(N)$.

To see that the solution $\varphi$ satisfies \emph{grand-coalition normalization}, let $v\in\mathcal{W}$. For each $i\neq n$, because $v(i)=0$, $\varphi_i\left(v^{v(N)-\sum_{k \in N}v(k)}\right)=x_i=\varphi_i(v)$. Moreover, note that $v(N)-\sum_{k \in N}v(k)=v(N)-v(n)$. Hence, $\varphi_n\left(v^{v(N)-\sum_{k \in N}v(k)}\right)=v(N)-v(n)=\varphi_n(v)-v(n)$. Note that for each $v\in\mathcal{V}\setminus\mathcal{W}$, $\varphi(v)=P(v)$. Because the proportional division value satisfies \emph{grand-coalition normalization}, so does $\varphi$.

\section{Composition Axioms}\label{sec:composition}

\subsection{Characterizations}\label{sec:composition_characterizations}
We now define TU-game versions of \emph{composition} axioms. These axioms were introduced in the context of bankruptcy problems, where a group of agents has claims on a resource that is insufficient to meet all claims, and the available resource must be allocated among them. \cite{young1988jet} and \cite{moulin2000econometrica} proposed composition axioms in this context.\footnote{Related composition ideas appear in several other settings. \cite{moulin1987ijgt} introduced a composition property in the context of surplus sharing problems. Composition axioms are also related to \emph{step-by-step negotiation}, introduced by \cite{kalai1977metrica} in bargaining problems, and to \emph{path independence} of choice functions, studied by \cite{plott1973metrica}.} Composition axioms formalize the idea that the final allocation should not depend on whether the problem is solved all at once or in stages.

In a bankruptcy problem, suppose that an allocation is first computed for a reference problem with a given amount of the resource. If the amount ultimately available is different, one can either solve the problem with the revised amount directly, keeping the original claims, or use the initial allocation to define a second problem and then solve this second problem. A composition axiom requires these two procedures to lead to the same final allocation. Depending on whether the revised amount of the resource is larger or smaller than the reference amount, this idea gives rise to two requirements, usually called \emph{composition up} and \emph{composition down}.

The first axiom, \emph{composition up}, considers cases in which the available amount of the resource is later found to be larger than the reference amount. The final allocation can then be determined in two ways. One can apply the rule directly to the problem with the revised amount, keeping the original claims. Alternatively, one can first apply the rule to the problem with the reference amount, subtract the resulting allocation from the original claims, and then apply the rule to the residual problem with the additional amount of the resource. The final allocation is obtained by adding the initial allocation and the allocation of the residual problem. \emph{Composition up} requires these two procedures to lead to the same final allocation.

The second axiom, \emph{composition down}, considers cases in which the available amount of the resource is later found to be smaller than the reference amount. The final allocation can then be determined in two ways. One can apply the rule directly to the problem with the revised amount, keeping the original claims. Alternatively, one can first apply the rule to the problem with the reference amount, treat the resulting allocation as the new claims, and then apply the rule to the problem with the revised amount of the resource and these new claims. The final allocation is the allocation obtained in this second-stage problem. \emph{Composition down} requires these two procedures to lead to the same final allocation.

Since TU-games provide a richer framework than bankruptcy problems, we adapt each composition axiom to the TU-game setting. In the bankruptcy interpretation, the object to be allocated is the available amount of the resource. In our TU-game formulation, the corresponding aggregate value is the grand coalition worth. Thus, the composition axioms below compare the solution applied to a target game with the outcome obtained by first considering a reference game, using the payoff vector assigned in that reference game, and then constructing an associated residual game. This formulation preserves the basic idea of composition while allowing for the richer structure of TU-games, where the worths of coalitions other than the grand coalition are also part of the model.

This richer domain also affects how the reference game is chosen. In bankruptcy problems, the resource is nonnegative, and the distinction between an increase and a decrease in the available amount is naturally expressed by an order relation. In general TU-games, by contrast, the grand coalition worths in the target and reference games may be positive, negative, or zero. Hence, an order restriction on the reference grand coalition worth, such as requiring it to be no larger than the target grand coalition worth, would not provide a domain-independent notion of an increase. We therefore allow the reference grand coalition worth to vary freely and impose only the requirement that the reference payoff vector and the induced residual game be well defined on the domain under consideration. In Section~\ref{sec:composition_characterizations}, we introduce restricted versions on a bankruptcy-type domain, where the usual order interpretation is available.

Let 
\[
\Upsilon^U := \left\{(x,v) \in \mathbb{R}^N \times \mathcal{V}: \sum_{k \in N} x_k \neq \sum_{k\in N}v(k)\right\}
\quad \text{and} \quad
\Upsilon^D := \left\{(x,v) \in \mathbb{R}^N \times \mathcal{V}: \sum_{k \in N} x_k \neq 0\right\}. 
\]

We first define the residual game associated with \emph{composition up}. Given a payoff vector $x$, the game $U(x,v)$ subtracts the payoff assigned to each member of a coalition from that coalition's worth. Thus, $U(x,v)$ represents the residual game obtained after the payoff vector $x$ has been treated as an initial allocation in the target game $v$. Formally, let $U: \Upsilon^U \to \mathcal{V}$ be such that, for each $S \subseteq N$,
\[
U(x,v)(S) := v(S) - \sum_{k \in S} x_k. 
\]

\noindent\textbf{Composition Up}: For each $v \in \mathcal{V}$ and each $\alpha \in \mathbb{R}$,
\[
\left(\varphi\left(v^\alpha\right),v\right) \in \Upsilon^U 
\ \Rightarrow \ 
\varphi(v) = \varphi\left(v^\alpha\right) 
+ \varphi\left(U\left(\varphi\left(v^\alpha\right),v\right)\right).
\]

In this axiom, $v$ is the target game and $v^\alpha$ is the reference game, obtained from $v$ by changing only the grand coalition worth to $\alpha$. Thus, $\alpha$ plays the role of a reference grand coalition worth. The vector $\varphi(v^\alpha)$ is the payoff vector assigned in the reference game. The residual game $U(\varphi(v^\alpha),v)$ is obtained by subtracting these reference payoffs from each coalition's worth in the target game. The axiom requires that solving the target game directly give the same payoff vector as first assigning the reference payoffs and then solving the residual game.

The reason we introduce $\Upsilon^U$ and $\Upsilon^D$ is to ensure that the residual games remain in the domain on which the solution is defined. This is important because the proportional division value is defined on $\mathcal{V}$. For example, for each $v \in \mathcal{V}$ and each $i \in N$, $P_i(v^{\sum_{k \in N}v(k)})=v(i)$. Hence, after assigning this payoff vector, the induced residual game has zero stand-alone worths for all players and therefore falls outside $\mathcal{V}$. Thus, without a domain restriction, the proportional division value would not be applicable to some residual games.

We next define the residual game associated with \emph{composition down}. Given a payoff vector $x$, the game $D(x,v)$ treats the components of $x$ as the new stand-alone worths of the players and assigns to each proper coalition the sum of its members' reference payoffs, while keeping the grand-coalition worth equal to $v(N)$. Formally, let $D: \Upsilon^D \to \mathcal{V}$ be such that, for each $S \subseteq N$,
\[
D(x,v)(S) :=
\begin{cases}
\sum_{k \in S} x_k &\text{if } S \neq N, \\
v(N) &\text{if } S = N. 
\end{cases}
\]

\noindent\textbf{Composition Down}: For each $v \in \mathcal{V}$ and each $\alpha \in \mathbb{R}$,
\[
\left(\varphi\left(v^\alpha\right),v\right) \in \Upsilon^D 
\ \Rightarrow \ 
\varphi(v) = \varphi\left(D\left(\varphi\left(v^\alpha\right),v\right)\right). 
\]

In this axiom, as in \emph{composition up}, $v$ is the target game and $v^\alpha$ is the reference game, obtained from $v$ by changing only the grand coalition worth to $\alpha$. Thus, $\alpha$ plays the role of a reference grand coalition worth. The vector $\varphi(v^\alpha)$ is the payoff vector assigned in the reference game. The transformed game $D(\varphi(v^\alpha),v)$ is obtained by assigning to each proper coalition the sum of the reference payoffs of its members, while keeping the grand coalition worth of the target game. The axiom requires that solving the target game directly give the same payoff vector as solving this transformed game.

\emph{Composition up} and \emph{composition down} therefore capture two ways in which a reference payoff vector can be used to construct a related game on the same player set. In \emph{composition up}, the reference payoffs are subtracted from coalition worths, and the solution to the residual game is added to the reference payoff vector. In \emph{composition down}, the reference payoffs define the worths of proper coalitions in the transformed game, and the solution is applied to this transformed game. In both cases, the player set is unchanged, and the axiom requires the payoff vector obtained directly from the target game to coincide with the payoff vector obtained through the reference game. Note that \emph{composition up} and \emph{composition down} are logically independent. However, once efficiency,\footnote{The definition is provided in Section~\ref{sec:nullifiedconsistency}.} linearity, and symmetry\footnote{This means that if two players are symmetric, their payoffs should be identical.} are imposed, either axiom implies the other.\footnote{One can also consider an alternative formulation of the composition down axiom. \citet{fknt2025wp} discuss the relationships among these variants in detail and prove the equivalence between \emph{composition up} and \emph{composition down} (see Section~4.1 and Theorem~2 of \citet{fknt2025wp}).}

Our next axiom concerns games in which the grand coalition creates no surplus or deficit beyond the players' stand-alone worths. In such games, the grand coalition worth is exactly equal to the sum of the values that the players can generate individually. Thus, although the game may still contain nontrivial intermediate coalition worths, the aggregate value available to all players is exhausted by their stand-alone worths. The axiom requires each player to receive the player's own stand-alone worth in this case.

The motivation is that, when cooperation by the grand coalition does not create any additional aggregate value, there is no surplus or deficit to redistribute at the grand coalition level. The stand-alone worths then provide a natural benchmark allocation: each player receives the value that the player can generate individually.

\medskip
\noindent\textbf{Strong Inessential-Game Property}: For each $v \in \mathcal{V}$,
\[
v(N) = \sum_{k \in N} v(k) \ \Rightarrow \ \text{for each } i \in N, \ \varphi_i(v) = v(i). 
\]

This requirement is related to the \emph{inessential game property} of \citet{rvz1996ijgt}. Their axiom imposes the same payoff requirement on additive games. Our version applies to the larger class of games in which the grand coalition worth equals the sum of stand-alone worths, even if intermediate coalition worths are not additive. For this reason, we call the axiom \emph{strong inessential-game property}.

\medskip

Finally, we define a weak continuity. 

\medskip
\noindent\textbf{Grand-Coalition Continuity}: $\varphi$ is continuous with respect to the grand coalition worth.

\medskip
We now present characterizations of the proportional division value derived from composition axioms.

\begin{theorem}\label{thm:composition}
On the domain $\mathcal{V}$, the following statements are equivalent. 
\begin{enumerate}\itemsep-0.1cm
\item[(i)] $\varphi = P$.
\item[(ii)] $\varphi$ satisfies \emph{composition up}, \emph{strong inessential-game property}, \emph{grand-coalition normalization}, and \emph{grand-coalition continuity}.
\item[(iii)] $\varphi$ satisfies \emph{composition down}, \emph{strong inessential-game property}, \emph{grand-coalition normalization}, and \emph{grand-coalition continuity}.
\end{enumerate}
\end{theorem}

Before proving the theorem, we describe the structure of the argument. The main task is to show that the axioms force the solution to be proportional to the vector of stand-alone worths as the grand-coalition worth varies. The proof proceeds through the following steps. Claim~\ref{claim:gn} shows that \emph{grand-coalition normalization} yields a recursive relation between the payoff in a game and the payoff in the game obtained by subtracting the aggregate stand-alone worth from the grand coalition worth. Claim~\ref{claim:integer} combines this relation with \emph{strong inessential-game property} to pin down payoffs at all integer multiples of the aggregate stand-alone worth. Claims~\ref{claim:n-1/n}, \ref{claim:1/n}, and~\ref{claim:m/n} then use \emph{composition up} to extend the conclusion from integer multiples to all rational multiples. Finally, \emph{grand-coalition continuity} extends the conclusion from rational multiples to all real multiples, and evaluating the resulting formula at the original grand coalition worth yields the proportional division value. The proof for \emph{composition down} follows the same overall strategy, but derives the rational-multiple step directly from \emph{composition down}.

\begin{proof}
(i) $\Rightarrow$ (ii) and (iii). The proportional division value trivially satisfies \emph{strong inessential-game property} and \emph{grand-coalition continuity}. In addition, Theorem~\ref{thm:homogeneity} shows that it also satisfies \emph{grand-coalition normalization}. Therefore, we now focus on \emph{composition up} and \emph{composition down}. Let $v \in \mathcal{V}$, $\alpha \in \mathbb{R}$, and $i \in N$. First, suppose that $\left(P\left(v^\alpha\right),v\right) \in \Upsilon^U$. 
\begin{eqnarray*}
P_i\left(v^\alpha\right) + P_i\left(U\left(P\left(v^\alpha\right),v\right)\right) &=& \frac{v(i)}{\sum_{k \in N}v(k)} \alpha + \frac{v(i)-\frac{v(i)}{\sum_{k \in N}v(k)} \alpha}{\sum_{k \in N} \left(v(k) - \frac{v(k)}{\sum_{k' \in N}v(k')} \alpha\right)} \left(v(N) - \alpha\right) \\
&=& \frac{v(i)}{\sum_{k \in N}v(k)} \alpha + \frac{\frac{\sum_{k \in N}v(k)-\alpha}{\sum_{k \in N}v(k)} v(i)}{\sum_{k \in N}v(k) - \alpha} \left(v(N) - \alpha\right) \\
&=& \frac{v(i)}{\sum_{k \in N}v(k)} \alpha + \frac{v(i)}{\sum_{k \in N}v(k)} \left(v(N) - \alpha\right) \\
&=& \frac{v(i)}{\sum_{k \in N}v(k)} v(N) = P_i(v). 
\end{eqnarray*}
Hence, the proportional division value satisfies \emph{composition up}. 

Now suppose that $\left(P\left(v^\alpha\right),v\right) \in \Upsilon^D$. 
\begin{eqnarray*}
P_i\left(D\left(P\left(v^\alpha\right),v\right)\right) &=& \frac{\frac{v(i)}{\sum_{k \in N}v(k)}\alpha}{\sum_{k \in N}\left(\frac{v(k)}{\sum_{k' \in N} v(k')}\alpha\right)}v(N) \\
&=& \frac{v(i)}{\sum_{k \in N}v(k)}\alpha \times \frac{1}{\frac{\sum_{k \in N}v(k)\alpha}{\sum_{k \in N}v(k)}} \times v(N)  \\
&=& \frac{v(i)}{\sum_{k \in N}v(k)}v(N) = P_i(v). 
\end{eqnarray*}
Hence, the proportional division value satisfies \emph{composition down}. 

\bigskip

\noindent (ii) $\Rightarrow$ (i). Let $\varphi$ be a solution satisfying \emph{grand-coalition normalization}. 

\begin{claim}\label{claim:gn}
For each $v \in \mathcal{V}$, each $i \in N$, and each $n \in \mathbb{Z}_{\geq 0}$,\footnote{We write $\mathbb{Z}_{\geq 0} = \{n \in \mathbb{Z}: n \geq 0\}$, and use analogous notations for other sets.}
\[
\varphi_i\left(v\right) = nv(i) + \varphi_i\left(v^{v(N) - n\sum_{k \in N} v(k)}\right). 
\]
\end{claim}

\noindent\emph{Proof of Claim~\ref{claim:gn}.}
Let $v \in \mathcal{V}$ and $i \in N$. By \emph{grand-coalition normalization}, 
\[
\varphi_i\left(v\right) = v(i) + \varphi_i\left(v^{v(N) - \sum_{k \in N} v(k)}\right). 
\]
Again by \emph{grand-coalition normalization},
\begin{eqnarray*}
\varphi_i\left(v^{v(N) - \sum_{k \in N} v(k)}\right) &=& v(i) + \varphi_i\left(v^{v(N) - \sum_{k \in N} v(k)-\sum_{k \in N} v(k)}\right) \\
&=& v(i) + \varphi_i\left(v^{v(N) - 2\sum_{k \in N} v(k)}\right). 
\end{eqnarray*}
Hence, 
\[
\varphi_i\left(v\right) = 2v(i) + \varphi_i\left(v^{v(N) - 2\sum_{k \in N} v(k)}\right).
\]
By applying an analogous argument repeatedly, we conclude that for each $n \in \mathbb{Z}_{\geq0}$,
\[
\varphi_i\left(v\right) = nv(i) + \varphi_i\left(v^{v(N) - n\sum_{k \in N} v(k)}\right). 
\]
\hfill $\blacksquare$

\medskip

Now in addition to \emph{grand-coalition normalization}, suppose that $\varphi$ satisfies \emph{strong inessential-game property}. 

\begin{claim}\label{claim:integer}
For each $v \in \mathcal{V}$, each $i \in N$, and each $n \in \mathbb{Z}$, 
\[
\varphi_i\left(v^{n \sum_{k \in N}v(k)}\right) = n v(i). 
\]
\end{claim}

\noindent\emph{Proof of Claim~\ref{claim:integer}.} 
Let $v \in \mathcal{V}$ and $i \in N$. First by \emph{strong inessential-game property},
\[
\varphi_i\left(v^{\sum_{k \in N} v(k)}\right) = v(i). 
\]
By Claim~\ref{claim:gn}, for each $m \in \mathbb{Z}_{>0}$, 
\begin{eqnarray*}
\varphi_i\left(v^{\sum_{k \in N} v(k)}\right) &=& m v(i) + \varphi_i\left(v^{(1-m)\sum_{k \in N} v(k)}\right). 
\end{eqnarray*}
Hence, 
\begin{eqnarray*}
v(i) = m v(i) + \varphi_i\left(v^{(1-m)\sum_{k \in N} v(k)}\right) \
\iff \ \varphi_i\left(v^{(1-m)\sum_{k \in N} v(k)}\right) = (1-m) v(i). 
\end{eqnarray*}
This implies that for each $n \in \mathbb{Z}_{\leq 0}$,
\[
\varphi_i\left(v^{n\sum_{k \in N} v(k)}\right) =n v(i). 
\]
Let $n \in \mathbb{Z}_{>1}$. By Claim~\ref{claim:gn}, 
\begin{eqnarray*}
\varphi_i\left(v^{n \sum_{k \in N} v(k)}\right) &=& (n-1)v(i) + \varphi_i\left(v^{(n-(n-1))\sum_{k \in N}v(k)}\right) \\
&=& (n-1)v(i) + v(i) \\
&=& nv(i). 
\end{eqnarray*}
\hfill$\blacksquare$

\medskip

Now suppose that $\varphi$ also satisfies \emph{composition up}. 

\begin{claim}\label{claim:n-1/n}
For each $v \in \mathcal{V}$, each $i \in N$, and each $n \in \mathbb{Z}_{\neq 0}$, 
\[
\varphi_i\left(v^{\frac{n-1}{n}\sum_{k \in N}v(k)}\right) = \frac{n-1}{n} v(i). 
\]
\end{claim}

\noindent\emph{Proof of Claim~\ref{claim:n-1/n}.}
Let $v \in \mathcal{V}$, $i \in N$, and $n \in \mathbb{Z}_{\neq 0}$. By Claim~\ref{claim:integer}, when $n = -1$ or $1$, the equation in the claim holds. Now suppose that $n  \notin \{-1,1\}$. Let $w \in \mathcal{V}$ be defined by
\[
w(S) = 
\begin{cases}
v(S) - \frac{n-1}{n} \sum_{k \in S} v(k) \ &\text{if} \ S \neq N, \\
\frac{n-1}{n} \sum_{k \in N} v(k) \ &\text{if} \ S = N. 
\end{cases}
\]
Note that for each $k \in N$, $w(k) = \frac{1}{n}v(k)$. By \emph{composition up}, and Claims~\ref{claim:gn} and~\ref{claim:integer},\footnote{Note that for each $j \in N$, $\varphi_j\left(w^{-(n-1)\sum_{k \in N} w(k)}\right) = -(n-1)w(j) = -\frac{n-1}{n}v(j)$. Because $n \neq 1$, we have $\sum_{k \in N} \varphi_k\left(w^{-(n-1)\sum_{k \in N} w(k)}\right) \neq 0$. This implies that $\left(\varphi\left(w^{-(n-1)\sum_{k \in N} w(k)}\right), w^0\right) \in \Upsilon^U$.} because $0 \in \mathbb{Z}$, 
\begin{eqnarray*}
0=\varphi_i\left(w^0\right) &=& \varphi_i\left(w^{-(n-1)\sum_{k \in N} w(k)}\right) + \varphi_i\left(U\left(\varphi\left(w^{-(n-1)\sum_{k \in N} w(k)}\right),w^0\right)\right), \\
 &=& -(n-1)w(i) + \varphi_i\left(U\left(\varphi\left(w^{-(n-1)\sum_{k \in N} w(k)}\right),w^0\right)\right). 
\end{eqnarray*}
Then, 
\begin{eqnarray*}
U\left(\varphi\left(w^{-(n-1)\sum_{k \in N} w(k)}\right),w^0\right)(S) 
&=&
\begin{cases}
w(S) - \left(-\sum_{k \in S} \varphi_k\left(w^{-(n-1)\sum_{k \in N} w(k)}\right)\right) \ &\text{if} \ S \neq N, \\
0 + \sum_{k \in N} (n-1) w(k) \ &\text{if} \ S = N, 
\end{cases}\\
&=&
\begin{cases}
w(S) + \sum_{k \in S} \left(n-1\right)w(k) \ &\text{if} \ S \neq N, \\
0 + \sum_{k \in N} (n-1) w(k) \ &\text{if} \ S = N, 
\end{cases}\\
&=&
\begin{cases}
v(S) - \frac{n-1}{n} \sum_{k \in S} v(k) + \sum_{k \in S} \frac{n-1}{n}v(k) \ &\text{if} \ S \neq N, \\
\sum_{k \in N} \frac{n-1}{n} v(k) \ &\text{if} \ S = N,
\end{cases} \\
&=&
\begin{cases}
v(S) \ &\text{if} \ S \neq N, \\
\frac{n-1}{n} \sum_{k \in N} v(k) \ &\text{if} \ S = N.
\end{cases}
\end{eqnarray*}
where the first equality follows from the definition of $U$, the second equality from Claim~\ref{claim:integer}, and the third equality from the definition of $w$. Specifically, since for each $k\in N$ we have
\[
w(k)=\frac{1}{n}v(k),
\]
it follows that for each $S\neq N$,
\begin{align*}
w(S)+\sum_{k\in S}(n-1)w(k)
&= \left(v(S)-\frac{n-1}{n}\sum_{k\in S}v(k)\right)
   +\sum_{k\in S}\left((n-1)\cdot\frac{1}{n}v(k)\right)\\
&= v(S)-\frac{n-1}{n}\sum_{k\in S}v(k)
   +\sum_{k\in S}\frac{n-1}{n}v(k).
\end{align*}
Consequently,
\[
\varphi_i\left(v^{\frac{n-1}{n} \sum_{k \in N} v(k)}\right) = (n-1)w(i) = \frac{n-1}{n}v(i). 
\]
\hfill $\blacksquare$

\medskip

\begin{claim}\label{claim:1/n}
For each $v \in \mathcal{V}$, each $n \in \mathbb{Z}_{\neq 0}$, and each $i \in N$,
\[
\varphi_i\left(v^{\frac{1}{n} \sum_{k \in N} v(k)}\right) = \frac{1}{n} v(i). 
\]
\end{claim}

\noindent\emph{Proof of Claim~\ref{claim:1/n}.}
Let $v \in \mathcal{V}$, $i \in N$, and $n \in \mathbb{Z}_{\neq  0}$. By \emph{grand-coalition normalization},
\begin{eqnarray*}
\frac{n-1}{n}v(i)=\varphi_i\left(v^{\frac{n-1}{n} \sum_{k \in N}v(k)}\right) &=& v(i) + \varphi_i\left(v^{\left(\frac{n-1}{n}-1\right)\sum_{k \in N} v(k)}\right) \\
&=& v(i) + \varphi_i\left(v^{-\frac{1}{n}\sum_{k \in N} v(k)}\right),
\end{eqnarray*}
and by Claim~\ref{claim:n-1/n},
\begin{eqnarray*}
\varphi_i\left(v^{-\frac{1}{n}\sum_{k \in N} v(k)}\right) &=& \frac{n-1}{n} v(i) - v(i) \\
&=& -\frac{1}{n} v(i). 
\end{eqnarray*}
Hence, by Claim~\ref{claim:n-1/n} applied to $-n$,
\[
\varphi_i\left(v^{\frac{1}{n}\sum_{k \in N} v(k)}\right) = \frac{1}{n}v(i).
\]
\hfill $\blacksquare$

\medskip

\begin{claim}\label{claim:m/n}
For each $v \in \mathcal{V}$, each $i \in N$, each $m \in \mathbb{Z}_{\geq 0}$, and each $n \in \mathbb{Z}_{\neq 0}$,
\[
\varphi_i\left(v^{\frac{m}{n} \sum_{k \in N}v(k)}\right) = \frac{m}{n} v(i). 
\]
\end{claim}

\noindent\emph{Proof of Claim~\ref{claim:m/n}.}
The proof proceeds by induction on $m$.
The statement holds for $m=0$ by Claim~\ref{claim:integer}, and for $m=1$ by Claim~\ref{claim:1/n}.
Now, let $m \ge 2$ and suppose that the statement holds up to $m-1$. Let $v \in \mathcal{V}$, $i \in N$ and $n \in \mathbb{Z}_{\neq 0}$.

\noindent\textbf{Case 1}: $n = m-1$. By \emph{grand-coalition normalization} and Claim~\ref{claim:1/n},
\begin{eqnarray*}
\varphi_i\left(v^{\frac{m}{n} \sum_{k \in N}v(k)}\right) &=& v(i) + \varphi_i\left(v^{\left(\frac{m}{n}-1\right)\sum_{k \in N} v(k)}\right) \\ 
&=& v(i) + \varphi_i\left(v^{\frac{1}{n}\sum_{k \in N} v(k)}\right) \\
&=& v(i) + \frac{1}{n} v(i) \\
&=& \frac{n+1}{n} v(i) = \frac{m}{n} v(i). 
\end{eqnarray*}

\noindent\textbf{Case 2}: $n \neq m-1$. By \emph{composition up},
\[
\varphi_i\left(v^{\frac{m}{n} \sum_{k \in N}v(k)}\right) = x_i+
\varphi_i
	\left(
		U\left(
			x , v^{\frac{m}{n} \sum_{k \in N}v(k)}
	\right)
\right),
\]
where $x=\varphi\left(v^{\frac{m-1}{n} \sum_{k \in N}v(k)}\right)$, and $x_i=\frac{m-1}{n} v(i)$ by the induction hypothesis.
Note that 
\begin{eqnarray*}
U\left(x, v^{\frac{m}{n} \sum_{k \in N}v(k)}\right)(S) &=& 
\begin{cases}
v(S) - \sum_{k \in S} \frac{m-1}{n}v(k) \ &\text{if} \ S \neq N, \\
\frac{m}{n} \sum_{k \in N}v(k) - \sum_{k \in N} \frac{m-1}{n} v(k) \ &\text{if} \ S = N, 
\end{cases} \\
&=& 
\begin{cases}
\frac{n-m+1}{n} v(S) \ &\text{if} \ |S| = 1, \\
v(S) - \sum_{k \in S} \frac{m-1}{n}v(k) \ &\text{if} \ 1 < |S| < |N|, \\
\frac{1}{n-m+1}\left(\frac{n-m+1}{n}\sum_{k \in N}v(k)\right) \ &\text{if} \ S = N. 
\end{cases}
\end{eqnarray*}
Then, we can apply Claim~\ref{claim:1/n} to the game $\frac{n-m+1}{n}v$ and obtain:
\[
\varphi_i\left(U\left(
x, v^{\frac{m}{n} \sum_{k \in N}v(k)}\right)\right) = \frac{1}{n-m+1} \left(\frac{n-m+1}{n}v(i)\right) = \frac{1}{n} v(i). 
\]
Overall, 
\[
\varphi_i\left(v^{\frac{m}{n} \sum_{k \in N}v(k)}\right) = \frac{m-1}{n} v(i) + \frac{1}{n} v(i) = \frac{m}{n} v(i). 
\]
\hfill $\blacksquare$

\medskip

Finally, in addition to \emph{strong inessential-game property}, \emph{grand-coalition normalization}, and \emph{composition up}, suppose that $\varphi$ satisfies \emph{grand-coalition continuity}. By \emph{grand-coalition continuity} and Claim~\ref{claim:m/n}, for each $v \in \mathcal{V}$, each $i \in N$, and each $\alpha \in \mathbb{R}$, 
\[
\varphi_i\left(v^{\alpha \sum_{k \in N}v(k)}\right) = \alpha v(i). 
\]
Let $\beta: \mathcal{V} \to \mathbb{R}$ be defined by 
\[
\beta(v) := \frac{v(N)}{\sum_{k \in N} v(k)}. 
\]
Then, for each $v \in \mathcal{V}$ and each $i \in N$,
\[
\varphi_i(v) = \varphi_i\left(v^{\beta(v)\sum_{k \in N} v(k)}\right) = \beta(v) \cdot v(i) = \frac{v(i)}{\sum_{k \in N}v(k)} v(N) = P_i(v). 
\]

\bigskip
\noindent (iii) $\Rightarrow$ (i). After Claim~\ref{claim:m/n}, the same argument as above applies, since the composition axioms are not invoked from that point onward. Hence, it remains only to prove the rational-multiple conclusion using \emph{composition down}. The purpose of the following argument is to obtain the same rational-multiple conclusion as Claim~\ref{claim:m/n}, but using \emph{composition down} rather than \emph{composition up}. Once this conclusion is obtained, \emph{grand-coalition continuity} yields the proportional division value exactly as in the previous part.

Let $\varphi$ be a solution satisfying \emph{composition down}, \emph{strong inessential-game property}, and \emph{grand-coalition normalization}. 
The proof proceeds by induction on $m$.
The statement holds for $m=0$ by Claim~\ref{claim:integer}.\footnote{Note that Claim~\ref{claim:integer} relies only on \emph{strong inessential-game property} and \emph{grand-coalition normalization}.}
Now, let $m \ge 1$ and suppose that the statement holds up to $m-1$.
Let $v \in \mathcal{V}$, $i \in N$ and $n \in \mathbb{Z}_{\neq 0}$. By \emph{composition down}, 
\begin{eqnarray*}
\varphi_i\left(v^{\frac{m}{n} \sum_{k \in N} v(k)}\right) &=& \varphi_i\left(D\left(\varphi\left(v^{\sum_{k \in N} v(k)}\right),v^{\frac{m}{n} \sum_{k \in N} v(k)}\right)\right).
\end{eqnarray*}
By Claim~\ref{claim:integer},
\[
\varphi_i\left(v^{\sum_{k \in N} v(k)}\right)=v(i). 
\]
This implies that 
\begin{eqnarray}
D\left(\varphi\left(v^{\sum_{k \in N} v(k)}\right),v^{\frac{m}{n} \sum_{k \in N} v(k)}\right)(S) &=& 
\begin{cases}
\sum_{k \in S} v(k) \ &\text{if} \ S \neq N, \\
\frac{m}{n}\sum_{k \in N}   v(k) \ &\text{if} \ S = N.
\end{cases}
\label{eq:D_mn}
\end{eqnarray}

Let $w \in \mathcal{V}$ be such that 
\[
w(S) = 
\begin{cases}
\frac{1}{n} \sum_{k \in S} v(k) \ &\text{if} \ S \neq N, \\
m \sum_{k \in N} w(k) \ &\text{if} \ S = N. 
\end{cases}
\]
By \emph{composition down},
\[
\varphi_i(w) =\varphi_i\left(D\left(\varphi\left(w^{n\sum_{k \in N} w(k)}\right),w\right)\right).
\]
By Claim~\ref{claim:integer} and definition of $w$,
$\varphi_i(w)=m w(i) = \frac{m}{n} v(i)$.
Hence,
\[
\frac{m}{n} v(i)=\varphi_i\left(D\left(\varphi\left(w^{n\sum_{k \in N} w(k)}\right),w\right)\right). 
\]
Again, by Claim~\ref{claim:integer},
\[
\varphi_i\left(w^{n\sum_{k \in N} w(k)}\right) = n w(i) = v(i). 
\]
Then,
\begin{eqnarray*}
D\left(\varphi\left(w^{n\sum_{k \in N} w(k)}\right),w\right)(S) &=& 
\begin{cases}
n\sum_{k \in S} w(k) \ &\text{if} \ S \neq N, \\
m \sum_{k \in N} w(k) \ &\text{if} \ S = N,
\end{cases} \\
&=& 
\begin{cases}
\sum_{k \in S} v(k) \ &\text{if} \ S \neq N, \\
\frac{m}{n} \sum_{k \in N} v(k) \ &\text{if} \ S = N. 
\end{cases}
\end{eqnarray*}
This is the same game as in~\eqref{eq:D_mn}.
Consequently,
\begin{eqnarray*}
\varphi_i\left(v^{\frac{m}{n} \sum_{k \in N} v(k)}\right) &=& \varphi_i\left(D\left(\varphi\left(v^{\sum_{k \in N} v(k)}\right),v^{\frac{m}{n} \sum_{k \in N} v(k)}\right)\right) \\
&=& \varphi_i\left(D\left(\varphi\left(w^{n\sum_{k \in N} w(k)}\right),w\right)\right) = \frac{m}{n} v(i). 
\end{eqnarray*}

Therefore, by \emph{grand-coalition continuity}, the same final argument as in the proof of $(ii)\Rightarrow(i)$ yields $\varphi=P$.
\end{proof}

We provide examples to demonstrate the independence of the axioms used in Theorem~\ref{thm:composition}. 
First, consider the CIS value. 
It satisfies \emph{composition up}, \emph{composition down},\footnote{For a proof that the CIS value satisfies \emph{composition up} and \emph{composition down}, see \cite{fknt2025wp}.} \emph{strong inessential-game property}, and \emph{grand-coalition continuity}. 
However, it violates \emph{grand-coalition normalization}. 
This follows from a more general incompatibility: any solution satisfying \emph{linearity} violates \emph{grand-coalition normalization}. 
Since the CIS value is \emph{linear}, it violates \emph{grand-coalition normalization}. 
We state this incompatibility result formally and provide a proof in Lemma~\ref{lemma:gn&l} in Appendix~\ref{appendix:incompatible}.

Second, we drop \emph{strong inessential-game property}. Let $\alpha \in \mathbb{R}_{>0}$. Define the solution $\varphi$ as follows: for each $v \in \mathcal{V}$ and each $i \in N$, set $\varphi_i(v) = P_i\left(v^{v(N)-\alpha}\right)$. Because the proportional division value satisfies \emph{composition up}, \emph{composition down}, and \emph{grand-coalition continuity}, so does $\varphi$. However, because the grand coalition of each $v \in \mathcal{V}$ is reduced by $\alpha$, $\varphi$ violates \emph{strong inessential-game property}. We prove that this solution satisfies \emph{grand-coalition normalization}. Let $v \in \mathcal{V}$ and $i \in N$. Then,
\begin{eqnarray*}
\varphi_i\left(v^{v(N)-\sum_{k \in N}v(k)}\right) &=& P_i\left(v^{v(N)-\alpha-\sum_{k \in N}v(k)}\right) \\
&=& \frac{v(i)}{\sum_{k \in N} v(k)} \left(v(N)-\alpha-\sum_{k \in N}v(k)\right) \\
&=& \frac{v(i)}{\sum_{k \in N} v(k)} \left(v(N)-\alpha\right)-\frac{\sum_{k \in N}v(k)}{\sum_{k \in N}v(k)}v(i) \\
&=& P_i\left(v^{v(N) - \alpha}\right) - v(i) = \varphi_i(v) - v(i). 
\end{eqnarray*}
Hence, $\varphi$ satisfies \emph{grand-coalition normalization}. 

Third, we drop \emph{composition up} or \emph{composition down}. Let $|N| = n$, and for notational convenience, label the players so that $N = \{1,\hdots,n\}$. Let $\eta: \mathbb{R} \to \mathbb{R}$ be defined by $\eta(\alpha) := \alpha + \sin(2 \pi \alpha)$. Note that $\eta(\alpha - 1) = \eta(\alpha) - 1$.  
Let $\mathcal{W} = \left\{v \in \mathcal{V}: v(1) = 1 \ \text{and for each } i \neq 1, \ v(i) = 0\right\}$. Define the solution $\varphi$ as follows: for each $v \in \mathcal{W}$, set $\varphi(v) = (\eta(v(N)),0,\hdots,0)$, and for each $v \in \mathcal{V} \setminus \mathcal{W}$, set $\varphi(v) = P(v)$. 

We first verify that $\varphi$ satisfies \emph{strong inessential-game property}. Consider $v \in \mathcal{V}$ such that $v(N)=\sum_{k\in N}v(k)$. If $v\in\mathcal{W}$, then $\sum_{k\in N}v(k)=1$, and hence $v(N)=1$. Therefore, $\varphi(v)=(\eta(1),0,\hdots,0)=(1,0,\hdots,0)$, which coincides with the vector of stand-alone worths. If $v\notin\mathcal{W}$, then $\varphi(v)=P(v)$, and since the proportional division value satisfies \emph{strong inessential-game property}, we again have $\varphi_i(v)=v(i)$ for each $i\in N$. Thus, $\varphi$ satisfies \emph{strong inessential-game property}. 

Since $\eta$ is continuous and the proportional division value is continuous with respect to the grand-coalition worth, $\varphi$ satisfies \emph{grand-coalition continuity}. On the other hand, there are $v \in \mathcal{W}$ and $\alpha \in \mathbb{R}$ such that $U\left(\varphi\left(v^\alpha\right),v\right)$ and $D\left(\varphi\left(v^\alpha\right),v\right)$ do not belong to $\mathcal{W}$. For such games, the solution switches from the non-proportional definition on $\mathcal{W}$ to the proportional division value outside $\mathcal{W}$, and the equalities required by \emph{composition up} and \emph{composition down} fail. Hence, $\varphi$ violates both \emph{composition up} and \emph{composition down}.

We next verify that $\varphi$ satisfies \emph{grand-coalition normalization}. Since $\varphi=P$ on $\mathcal{V}\setminus\mathcal{W}$ and the proportional division value satisfies \emph{grand-coalition normalization}, it remains to check games in $\mathcal{W}$. Moreover, if $v\in\mathcal{W}$, then $v^{\,v(N)-\sum_{k \in N} v(k)} \in \mathcal{W}$. Let $v \in \mathcal{W}$. For each $i \neq 1$, $\varphi_i\left(v^{v(N)-\sum_{k \in N}v(k)}\right) = 0 =\varphi_i(v) - v(i)$. For player $1$,
\begin{eqnarray*}
\varphi_1\left(v^{v(N)-\sum_{k \in N}v(k)}\right) &=& \varphi_1\left(v^{v(N)-v(1)}\right) \\
&=& \varphi_1\left(v^{v(N)-1}\right)\\
&=& \eta\left(v(N) -1\right) \\
&=& \eta\left(v(N)\right) - 1 \\
&=& \varphi_1(v) - v(1). 
\end{eqnarray*}
Hence, $\varphi$ satisfies \emph{grand-coalition normalization}. 
 Let $v \in \mathcal{W}$. For each $i \neq 1$, 
$$
\varphi_i\left(v^{v(N)-\sum_{k \in N}v(k)}\right) = 0 =\varphi_i(v) - v(i). 
$$
Now,
\begin{eqnarray*}
\varphi_1\left(v^{v(N)-\sum_{k \in N}v(k)}\right) &=& \varphi_1\left(v^{v(N)-v(1)}\right) \\
&=& \varphi_1\left(v^{v(N)-1}\right)\\
&=& \eta\left(v(N) -1\right) \\
&=& \eta\left(v(N)\right) - 1 \\
&=& \varphi_1(v) - v(1). 
\end{eqnarray*}
Hence, $\varphi$ satisfies \emph{grand-coalition normalization}. 

Finally we drop \emph{grand-coalition continuity}. Let $\alpha \in \mathbb{Q}_{>0}$. Define the solution $\varphi$ as follows: for each $v \in \mathcal{V}$, if for each $i \in N$, $v(i) \in \mathbb{Z}$, and $v(N) \notin \mathbb{Q}$, then set $\varphi(v) = P\left(v^{v(N) - \alpha}\right)$; otherwise, set $\varphi(v) = P(v)$. Because the proportional division value satisfies \emph{composition up}, \emph{composition down}, \emph{strong inessential-game property}, and \emph{grand-coalition normalization}, so does $\varphi$. On the other hand, the solution clearly violates \emph{grand-coalition continuity}.

\subsection{Connection to Composition Axioms in Bankruptcy Problems}\label{subsec:bankruptcy_connection}
The bankruptcy literature contains several composition-based characterizations. 
\citet{young1988jet} characterizes the proportional solution using a composition axiom and \emph{self-duality}.\footnote{We use the term ``proportional solution'' rather than ``proportional division value'' in the context of bankruptcy problems. An alternative proof of this characterization is provided by \citet{thomson2016el}. More precisely, \citet{thomson2016el} proves the characterization of the proportional solution based on \emph{composition down}; the corresponding characterization based on \emph{composition up} follows by duality.}
More broadly, \citet{moulin2000econometrica}, \citet{chambers2006geb}, and \citet{chambersmorenoternero2017scw} study how composition axioms restrict rules. These results provide a benchmark for our analysis: we do not directly translate the bankruptcy axioms, but rather adapt their direct-versus-staged allocation logic to TU-games, where the characteristic function contains coalition values beyond the grand coalition worth.

To make the connection with bankruptcy problems precise, we restrict attention to a bankruptcy-type subclass of TU-games,
\[
\mathcal{V}^{bank}
:=
\left\{
v\in\mathcal{V}:
\text{for each } i\in N,\ 0\le v(i)
\ \text{and}\ 
0\le v(N)\le \sum_{k\in N} v(k)
\right\}.
\]
By construction, $\mathcal{V}^{bank}\subseteq\mathcal{V}$; hence each $v\in\mathcal{V}^{bank}$ satisfies $\sum_{k\in N}v(k)\neq 0$. Together with the non-negativity of singleton worths, this implies $\sum_{k\in N}v(k)>0$. This restriction mirrors the basic structure of bankruptcy problems: each agent's claim is non-negative, and the available resource does not exceed the sum of claims. In this subclass, the singleton worths play the role of claims, while the grand coalition worth plays the role of the available resource. Within $\mathcal{V}^{bank}$, we provide characterizations of the proportional division value based on composition axioms.

We reformulate the composition axioms so that they are invoked only when the residual game remains in $\mathcal{V}^{bank}$.\footnote{In the bankruptcy setting, feasibility is built into the definition of a solution: each agent's award must lie between $0$ and the agent's claim, and the endowment is fully allocated. As a result, the residual problem automatically remains within the bankruptcy domain. In contrast, in the TU-game setting we do not impose such feasibility restrictions a priori, so we explicitly require that the residual game remain in $\mathcal{V}^{bank}$.} To keep these reformulations close to the bankruptcy setting, we also restrict the reference grand coalition worth $\alpha$ to the natural feasible range. For \emph{limited composition up}, the reference worth satisfies $\alpha\in[0,v(N)]$, so the target game has at least as much aggregate value as the reference game. For \emph{limited composition down}, the reference worth satisfies $\alpha\in\left[v(N),\sum_{k\in N}v(k)\right]$, so the reference game has at least as much aggregate value as the target game but remains within the bankruptcy-type range. Under these restrictions, \emph{limited composition up} and \emph{limited composition down} parallel the corresponding composition axioms in the bankruptcy setting.

\medskip

\noindent\textbf{Limited Composition Up}: 
For each $v \in \mathcal{V}^{bank}$, each $\alpha \in [0,v(N)]$, and each $i \in N$,
\[
\left(\varphi\left(v^\alpha\right),v\right) \in \Upsilon^U
\ \text{and}\ 
U\left(\varphi\left(v^\alpha\right),v\right) \in \mathcal{V}^{bank}
\ \Rightarrow\ 
\varphi_i(v)
=
\varphi_i\left(v^\alpha\right)
+
\varphi_i\left(U(\varphi(v^\alpha),v)\right).
\]

\medskip
\noindent\textbf{Limited Composition Down}: 
For each $v \in \mathcal{V}^{bank}$, each $\alpha \in \left[v(N),\sum_{k \in N}v(k)\right]$, and each $i \in N$,
\[
\left(\varphi\left(v^\alpha\right),v\right) \in \Upsilon^D
\ \text{and}\ 
D\left(\varphi\left(v^\alpha\right),v\right) \in \mathcal{V}^{bank}
\ \Rightarrow\ 
\varphi_i(v)
=
\varphi_i\left(D\left(\varphi\left(v^\alpha\right),v\right)\right).
\]

Thus, on $\mathcal{V}^{bank}$, the limited composition axioms recover the order-based interpretation of composition axioms in bankruptcy problems, while the unrestricted axioms introduced earlier retain the flexibility needed for general TU-games.

To further motivate our focus on $\mathcal{V}^{bank}$, we describe a bankruptcy-to-game construction that embeds bankruptcy problems into TU-games and yields games in $\mathcal{V}^{bank}$. 
Since bankruptcy problems can be mapped to TU-games in more than one way, we consider the following construction, which specifies the singleton values and the grand coalition worth, while leaving other coalition worths unrestricted.

Given $(c,E)\in \mathbb{R}_+^N\times \mathbb{R}_+$ with $\sum_{k\in N} c_k \ge E$, let $\mathcal{V}^{(c,E)}\subseteq \mathcal{V}^{all}$ be the set of TU-games $v$ such that
\[
v(i)=c_i \ \text{for each } i\in N,
\qquad
v(N)=E,
\]
and no restriction is imposed on $v(S)$ for coalitions $S\subsetneq N$ with $2\le |S|$. 
This construction recovers $\mathcal{V}^{bank}$ in the following sense: a TU-game $v$ belongs to $\mathcal{V}^{bank}$ if and only if there is a bankruptcy problem $(c,E)\in \mathbb{R}_+^N\times \mathbb{R}_+$ with $\sum_{k\in N} c_k \ge E$ such that $v\in \mathcal{V}^{(c,E)}$.
We thus focus on the domain $\mathcal{V}^{bank}$, using this correspondence. Further discussion on other constructions is in Appendix \ref{sec:btoU}.

\medskip

Using \emph{limited composition up} and \emph{limited composition down}, we establish characterizations of the proportional division value that are parallel to the bankruptcy characterizations of \citet{young1988jet} and \citet{thomson2016el}. In bankruptcy problems, the proportional solution can be characterized by a composition axiom together with \emph{self-duality}. \emph{Self-duality} requires that, for each bankruptcy problem, the rule coincide with its recommendation for the dual problem.

Our result should be understood as a TU-game analogue of this characterization, rather than as a direct translation. In bankruptcy problems, the claims vector and the available resource are the relevant primitives. In the present TU-game setting, by contrast, the characteristic function contains coalition values beyond the singleton worths and the grand coalition worth. Thus, additional conditions are needed to control how the solution behaves on this richer domain. In our formulation, \emph{reverse grand-coalition normalization} plays the role analogous to \emph{self-duality}, while \emph{strong inessential-game property} and \emph{grand-coalition continuity} ensure that the characterization extends from the bankruptcy-type benchmark to the TU-game setting.

We now introduce \emph{reverse grand-coalition normalization}, which can be viewed as a reverse counterpart of \emph{grand-coalition normalization}. The reverse transformation is natural on $\mathcal{V}^{bank}$ because it preserves the bankruptcy-type domain. Indeed, on this domain, the transformation used in \emph{grand-coalition normalization} changes the grand-coalition worth to $v(N)-\sum_{k\in N}v(k)$, which is non-positive and remains in $\mathcal{V}^{bank}$ only in the boundary case $v(N)=\sum_{k\in N}v(k)$. By contrast, the reverse transformation changes the grand coalition worth to $\sum_{k\in N}v(k)-v(N)$, which remains in the bankruptcy-type range. Thus, \emph{reverse grand-coalition normalization} is the domain-preserving analogue of \emph{self-duality} in this setting.

\medskip
\noindent\textbf{Reverse Grand-Coalition Normalization}: 
For each $v \in \mathcal{V}^{bank}$ and each $i \in N$,
\[
\varphi_i\left(v^{\sum_{k \in N} v(k) - v(N)}\right)
=
v(i) - \varphi_i(v).
\]

\medskip

Based on these properties, we obtain the following equivalences.

\begin{theorem}\label{thm:young}
On the domain $\mathcal{V}^{bank}$, the following statements are equivalent.
\begin{enumerate}\itemsep-0.1cm
\item[(i)] $\varphi = P$. 
\item[(ii)] $\varphi$ satisfies \emph{limited composition up}, \emph{reverse grand-coalition normalization}, \emph{strong inessential-game property}, and \emph{grand-coalition continuity}. 
\item[(iii)] $\varphi$ satisfies \emph{limited composition down}, \emph{reverse grand-coalition normalization}, \emph{strong inessential-game property}, and \emph{grand-coalition continuity}. 
\end{enumerate}
\end{theorem}

Before proving the theorem, we describe the structure of the argument. The proof adapts the logic of the bankruptcy characterization to the bankruptcy-type domain $\mathcal{V}^{bank}$. The key step is to show that the axioms pin down the solution at all dyadic multiples of the aggregate stand-alone worth. First, \emph{strong inessential-game property} and \emph{reverse grand-coalition normalization} determine the solution at the endpoints and at the midpoint: the grand coalition worths $0$, $\frac{1}{2}\sum_{k\in N}v(k)$, and $\sum_{k\in N}v(k)$. Then \emph{limited composition up} extends this conclusion inductively to all dyadic points in $[0,1]$. Finally, \emph{grand-coalition continuity} extends the conclusion from dyadic points to all points in $[0,1]$, and evaluating the resulting formula at the original grand coalition worth yields the proportional division value. The proof based on \emph{limited composition down} follows the same logic and is given in Appendix~\ref{appendix:LCD}.

\begin{proof}
(i) $\Rightarrow$ (ii) and (iii). 
We only verify that the proportional division value satisfies \emph{reverse grand-coalition normalization}. 
Let $v \in \mathcal{V}^{bank}$ and $i \in N$. Then
\[
P_i\left(v^{\sum_{k \in N}v(k) - v(N)}\right) 
= 
\frac{v(i)}{\sum_{k \in N}v(k)}
\left(\sum_{k \in N}v(k) - v(N)\right)
=
v(i) - P_i(v).
\]

\medskip
\noindent (ii) $\Rightarrow$ (i).
Let $\varphi$ be a solution satisfying \emph{strong inessential-game property}, \emph{reverse grand-coalition normalization}, and \emph{limited composition up}. 

\begin{claim}\label{claim:lcu_101/2}
For each $v \in \mathcal{V}^{bank}$ and each $i \in N$,
\[
\varphi_i\left(v^{\sum_{k \in N} v(k)}\right) = v(i), 
\qquad
\varphi_i\left(v^0\right) = 0,
\qquad\text{and}\qquad
\varphi_i\left(v^{\frac{1}{2}\sum_{k \in N}v(k)}\right) = \frac{1}{2} v(i).
\]
\end{claim}

\noindent\emph{Proof of Claim~\ref{claim:lcu_101/2}.}
Let $v \in \mathcal{V}^{bank}$ and $i \in N$. 
By \emph{strong inessential-game property}, $\varphi_i\left(v^{\sum_{k \in N} v(k)}\right) = v(i)$. 
Applying \emph{reverse grand-coalition normalization} with $v(N)=\sum_{k\in N}v(k)$ yields
\[
\varphi_i\left(v^{\sum_{k \in N}v(k)}\right) 
=
v(i) - \varphi_i\left(v^{0}\right),
\]
and hence $\varphi_i(v^{0})=0$. 
Applying \emph{reverse grand-coalition normalization} again gives
\[
\varphi_i\left(v^{\frac{1}{2}\sum_{k \in N}v(k)}\right)
=
v(i) - \varphi_i\left(v^{\frac{1}{2}\sum_{k \in N}v(k)}\right),
\]
so $\varphi_i\left(v^{\frac{1}{2}\sum_{k \in N}v(k)}\right)=\frac{1}{2}v(i)$.
\hfill$\blacksquare$

\medskip

Let $S_0 := \{0,1\}$. 
For each $n \in \mathbb{Z}_{>0}$, define
\[
S_n 
:= 
S_{n-1} 
\cup 
\left\{\frac{a+b}{2}:\ a \text{ and } b \text{ are consecutive points in } S_{n-1}\right\}.
\]

\begin{claim}\label{claim:lcu_induction}
For each $v \in \mathcal{V}^{bank}$, each $i \in N$, each $n \in \mathbb{Z}_{\geq 0}$, and each $m \in S_n$,
\[
\varphi_i\left(v^{m\sum_{k \in N}v(k)}\right) = mv(i).
\]
\end{claim}

\noindent\emph{Proof of Claim~\ref{claim:lcu_induction}.}
The proof proceeds by induction on $n$. 
By Claim~\ref{claim:lcu_101/2}, the equality holds for $m\in S_0=\{0,1\}$. 
Assume it holds for every $m\in S_{n-1}$ with $n\ge 1$. 
Let $m\in S_n\setminus S_{n-1}$, $v\in\mathcal{V}^{bank}$, and $i\in N$.

\smallskip
\noindent\textbf{Case 1:} $m \ge \frac{1}{2}$. 
By Claim~\ref{claim:lcu_101/2}, the equality holds when $m=\frac{1}{2}$. 
Assume now that $m>\frac{1}{2}$ and let $m' := m-\frac{1}{2}$. 
By \emph{limited composition up},
\[
\varphi_i\left(v^{m\sum_{k\in N}v(k)}\right)
=
\varphi_i\left(v^{\frac{1}{2}\sum_{k\in N}v(k)}\right)
+
\varphi_i\left(
U\left(\varphi\left(v^{\frac{1}{2}\sum_{k\in N}v(k)}\right),\,v\right)
\right).
\]
By Claim~\ref{claim:lcu_101/2},
\[
\varphi_i\left(v^{\frac{1}{2}\sum_{k\in N}v(k)}\right)=\frac{1}{2}v(i).
\]
Moreover, for each $k\in N$,
\[
U\left(\varphi\left(v^{\frac{1}{2}\sum_{k\in N}v(k)}\right),\,v\right)(k)=\frac{1}{2}v(k),
\]
and
\[
U\left(\varphi\left(v^{\frac{1}{2}\sum_{k\in N}v(k)}\right),\,v\right)(N)
=
m'\sum_{k\in N}v(k)
=
2m'\sum_{k\in N}
U\left(\varphi\left(v^{\frac{1}{2}\sum_{k\in N}v(k)}\right),\,v\right)(k).
\]
Thus the residual game lies in $\mathcal{V}^{bank}$, and since $m\in S_n$, we have $2m'\in S_{n-1}$. 
By the induction hypothesis,
\[
\varphi_i\left(
U\left(\varphi\left(v^{\frac{1}{2}\sum_{k\in N}v(k)}\right),\,v\right)
\right)
=
2m'\cdot \frac{1}{2}v(i)
=
m'v(i).
\]
Therefore,
\[
\varphi_i\left(v^{m\sum_{k\in N}v(k)}\right)
=
\frac{1}{2}v(i)+m'v(i)
=
mv(i).
\]

\smallskip
\noindent\textbf{Case 2:} $m < \frac{1}{2}$. 
By \emph{reverse grand-coalition normalization},
\[
\varphi_i\left(v^{m\sum_{k \in N}v(k)}\right)
=
v(i) - \varphi_i\left(v^{(1-m)\sum_{k \in N}v(k)}\right).
\]
Since $1-m>\frac{1}{2}$, Case~1 implies
$\varphi_i\left(v^{(1-m)\sum_{k \in N}v(k)}\right)=(1-m)v(i)$, and hence
\[
\varphi_i\left(v^{m\sum_{k \in N}v(k)}\right)
=
v(i)-(1-m)v(i)
=
mv(i).
\]
\hfill$\blacksquare$

\medskip

Finally, assume that $\varphi$ also satisfies \emph{grand-coalition continuity}. 
Since $S_n$ is dense in $[0,1]$ and $\alpha\mapsto v^{\alpha\sum_{k\in N}v(k)}$ varies continuously in~$\alpha$, \emph{grand-coalition continuity} implies that
\[
\varphi_i\left(v^{\alpha\sum_{k \in N} v(k)}\right) = \alpha v(i)
\qquad
\text{for all } \alpha\in[0,1].
\]
In particular, taking $\alpha = \frac{v(N)}{\sum_{k\in N}v(k)}$ yields $\varphi_i(v)=P_i(v)$, and hence $\varphi=P$.

\medskip
\noindent (iii) $\Rightarrow$ (i).
The proof follows the same strategy as the proof of $(ii)\Rightarrow(i)$. The only difference is that the dyadic induction step is obtained from \emph{limited composition down} rather than from \emph{limited composition up}. We provide the details in Appendix~\ref{appendix:LCD}.
\end{proof}

\section{A Nullified-Game Consistency Axiom}\label{sec:nullifiedconsistency}
We define a \emph{nullified-game consistency} axiom, proposed by \citet{kn2025el}.\footnote{
They name the axiom \emph{F-nullified reduced game consistency}, which is formulated for multi-valued solutions (or correspondences). This axiom builds on \emph{projection consistency}, originally proposed by \citet{funaki1996wp} and \citet{fy2001igtr}, which is a variable-population consistency axiom. For a comprehensive discussion of the concepts, justifications, and relationships between nullified-game consistencies, including other variants, and variable-population consistencies, we refer to \citet{kn2025el}.
}
The axiom is also related to \emph{partial implementation invariance} in bankruptcy problems, introduced by \citet{dtt2024scw}. In that setting, some agents' awards are treated as already implemented; the problem is then adjusted to account for these fixed awards, and the rule is required to preserve the awards of the remaining agents. Our nullified-game consistency axiom adapts this fixed-population logic to TU-games. The set of players remains fixed, but the game is modified so that the players whose payoffs have been fixed are treated as null players in the residual game.

Given $v \in \mathcal{V}$, we say that player $i \in N$ is a \emph{null} player in $v$ if for each $\emptyset \neq S \subseteq N$, $v(S \cup \{i\})-v(S)=0$. 
The axiom requires that when the recommended payoffs of some players are fixed and these players are hypothetically considered as null players, the solution restricted to the remaining players coincides with their original payoffs. In this setting, the population of players is unchanged, but the ``nullified'' players, whose payoffs have been fixed, no longer contribute to the residual worth of coalitions, since their shares are already settled. The axiom therefore requires that the solution applied to the residual game, in which the contributions of the nullified players are neutralized, reproduces the original payoffs for the remaining players. This captures the idea that the payoffs among the remaining players depend only on the residual worth available to them, independently of the fixed payoffs of the nullified players.

In this section, we assume $|N| \geq 3$ to exclude trivial cases. Moreover, to guarantee the existence of residual games, we restrict attention to the subclass 
$\widehat{\mathcal{V}} := \left\{v \in \mathcal{V}:
\text{either for each} \ i \in N, \ v(i) \geq 0, \
\text{or for each} \ i \in N, \ v(i) \leq 0
\right\}$. 

Given $S \subseteq N$, let $R^S: \mathbb{R}^N \cdot \widehat{\mathcal{V}} \to \mathcal{V}^{all}$ be defined by 
$$
R^S(x,v)(T) :=
\begin{cases}
v(N) - \sum_{k \in N \setminus S} x_k \ &\text{if} \ S \subseteq T \\
v(T \cap S) \ &\text{otherwise}. 
\end{cases}
$$
By definition, note that each player $i \notin S$ is a null player in $R^S(x,v)$.

\medskip
\noindent\textbf{Projection Nullified-Game Consistency}: For each $v \in \widehat{\mathcal{V}}$, each $S \subseteq N$ such that there is $j \in S$ with $v(j) \neq 0$, and each $i \in S$, 
\[
\varphi_i(v)=\varphi_i\left(R^S\left(\varphi(v),v\right)\right).
\]

Next, we consider the standard efficiency axiom, which requires that the total payoffs assigned to all players exactly equal the grand coalition worth.

\medskip
\noindent\textbf{Efficiency}: For each $v \in \widehat{\mathcal{V}}$,
\[
\sum_{k \in N} \varphi_k(v) = v(N). 
\]

The final axiom, \emph{equal ratio for two players}, specifies how payoffs are determined in simple two-player environments. 
Recall that a player is considered a \emph{null} player if their presence does not change the worth of any coalition. \emph{Equal ratio for two players} requires that, when all other players are null, the ratio of each active player’s payoff to the stand-alone worth of the other player is equal. In other words, the payoffs between two non-null players should be proportional to their individual stand-alone worth. This axiom embodies a fairness principle in minimal cooperative settings: when only two players generate any surplus, their payoffs reflect the relative scale of their contributions, ensuring that no player is favored disproportionately and that the division respects the relative worth of the players involved.

\medskip
\noindent\textbf{Equal Ratio for Two Players}: For each $v \in \widehat{\mathcal{V}}$ and each pair $i,j \in N$, 
\[
\text{each} \ k \neq i, j \ \text{is a null player in} \ v \ \Rightarrow \ \varphi_i(v)  \cdot v(j)=\varphi_j(v)  \cdot v(i).
\]

\cite{zvcf2021scw} define a similar axiom, which they call \emph{proportional-balanced treatment}, imposing the same requirement on games in which either all stand-alone worths are positive or all stand-alone worths are negative. They provide a characterization of the proportional division value for the subclass of two-player TU-games using proportional-balanced treatment together with \emph{efficiency}.

It is natural to consider \emph{projection nullified-game consistency} and \emph{equal ratio for two players} together, as they capture complementary aspects of fairness and consistency in cooperative games. \emph{Projection nullified-game consistency} ensures that payoffs among a subset of players remain unchanged when other players’ payoffs are fixed, while \emph{equal ratio for two players} imposes proportionality in the simplest two-player setting. Together, they express the principle that the rule should handle residual worth systematically and treat active players’ payoffs in a manner that reflects their individual contributions.

Here is our characterization of the proportional division value based on \emph{projection nullified-game consistency}. 
\begin{theorem}\label{thm:nullifiedconsistencyer}
A solution $\varphi: \widehat{\mathcal{V}} \to \mathbb{R}^N$ satisfies \emph{projection nullified-game consistency}, \emph{efficiency}, and \emph{equal ratio for two players} if and only if $\varphi = P$. 
\end{theorem}
\begin{proof}
$(\Leftarrow)$ We prove that the proportional division value satisfies \emph{projection nullified-game consistency}. Let $v \in \widehat{\mathcal{V}}$ and $S \subseteq N$ be such that there is $j \in S$ with $v(j) \neq 0$.
By definition, 
\begin{eqnarray*}
R^S(P(v),v)(T) 
&=& 
\begin{cases}
v(N) - \sum_{k \in N \setminus S} \frac{v(k)}{\sum_{k' \in N}v(k')}v(N) \ &\text{if} \ S \subseteq T, \\
v(T \cap S) \ &\text{otherwise}, 
\end{cases} \\
&=& 
\begin{cases}
\sum_{k \in S} \frac{v(k)}{\sum_{k' \in N}v(k')}v(N) \ &\text{if} \ S \subseteq T, \\
v(T \cap S) \ &\text{otherwise}.    
\end{cases}
\end{eqnarray*}
Then for each $i \in S$: 
\begin{enumerate}\itemsep-0.1cm
\item If $S=\{i\}$, then for each $k\neq i$, we have $\{k\}\cap\{i\}=\emptyset$, which implies $R^{S}= \left(P(v),v\right)(k)=0$.
Therefore,
$$
P_i\left(R^S\left(P(v),v\right)\right) = \frac{\frac{v(i)}{\sum_{k \in N}v(k)}v(N)}{\frac{v(i)}{\sum_{k \in N}v(k)}v(N) + \sum_{k \neq i} 0}\left(\frac{v(i)}{\sum_{k \in N}v(k)}v(N)\right) = \frac{v(i)}{\sum_{k \in N}v(k)}v(N) = P_i(v).
$$
\item If $\{i\} \subsetneq S$, 
\end{enumerate}
\begin{eqnarray*}
P_i\left(R^S\left(P(v),v\right)\right) &=& \frac{v(i)}{\sum_{k \in S} v(k)} \left(\sum_{k \in S} \frac{v(k)}{\sum_{k' \in N} v(k')} v(N)\right) \\
&=& \frac{v(i)}{\sum_{k \in S} v(k)} \left(\frac{\sum_{k \in S} v(k)}{\sum_{k' \in N} v(k')} v(N)\right) \\
&=& \frac{v(i)}{\sum_{k' \in N} v(k')} v(N) = P_i(v). 
\end{eqnarray*}
Hence, the proportional division value satisfies \emph{projection nullified-game consistency}. 

For each $v \in \widehat{\mathcal{V}}$ and each pair $i,j \in N$,
\begin{eqnarray*}
P_i(v) \cdot v(j) &=& 
\frac{v(i)}{\sum_{k \in N} v(k)} v(N) \cdot v(j) \\
&=& \frac{v(j)}{\sum_{k \in N} v(k)} v(N) \cdot v(i) \\
&=& P_j(v) \cdot v(i).
\end{eqnarray*}
This implies that the proportional division value satisfies \emph{equal ratio for two players}. 

\bigskip

\noindent $(\Rightarrow)$ Let $\varphi$ be a solution on $\widehat{\mathcal{V}}$ satisfying \emph{projection nullified-game consistency}, \emph{efficiency}, and \emph{equal ratio for two players}. 

\begin{claim}\label{claim:nullifiedconsistency}
For each $v \in \widehat{\mathcal{V}}$ and each pair $i,j \in N$, 
$$
\varphi_i(v) \cdot v(j) = \varphi_j(v) \cdot v(i). 
$$
\end{claim}

\noindent\emph{Proof of Claim \ref{claim:nullifiedconsistency}.}
Let $v \in \widehat{\mathcal{V}}$ and $i,j \in N$. \\
\textbf{Case 1}: $v(i) + v(j) = 0$. This implies that $v(i) = v(j) = 0$. Then clearly, 
$\varphi_i(v) \cdot v(j) = \varphi_j(v) \cdot v(i)$. 

\noindent\textbf{Case 2}: $v(i) + v(j) \neq 0$. This implies that (at least) either $v(i) \neq 0$ or $v(j) \neq 0$. By construction, each $k \neq i,j$ is a null player in $R^{\{i,j\}}(\varphi(v),v)$.\footnote{Note that when $i = j$, $R^{\{i,j\}}(\varphi(v),v) = R^{\{i,i\}}(\varphi(v),v) = R^{\{i\}}(\varphi(v),v)$.} Then by \emph{projection nullified-game consistency} and \emph{equal ratio for two players},
\begin{eqnarray*}
\varphi_i(v) \cdot v(j)-\varphi_j(v) \cdot v(i) &=& \varphi_i\left(R^{\{i,j\}}\left(\varphi(v),v\right)\right)\cdot v(j)-\varphi_j\left(R^{\{i,j\}}\left(\varphi(v),v\right)\right) \cdot v(i)\\
&=& \varphi_i\left(R^{\{i,j\}}\left(\varphi(v),v\right)\right)\cdot R^{\{i,j\}}\left(\varphi(v),v\right)(j)-\varphi_j\left(R^{\{i,j\}}\left(\varphi(v),v\right)\right) \cdot R^{\{i,j\}}\left(\varphi(v),v\right)(i)\\
&=& 0.    
\end{eqnarray*}
\hfill$\blacksquare$

\medskip

By summing over all agents on both sides of the equation in Claim \ref{claim:nullifiedconsistency}, 
\begin{eqnarray*}
\sum_{k \in N} \left(\varphi_i(v) \cdot v(k)\right) = \sum_{k \in N} \left(\varphi_k(v) \cdot v(i)\right) \ 
\iff \ \varphi_i(v) \cdot \sum_{k \in N} v(k) = v(i) \cdot \sum_{k \in N} \varphi_k(v). 
\end{eqnarray*}
By \emph{efficiency}, 
$
\sum_{k \in N} \varphi_k(v) = v(N), 
$
and thus 
\begin{eqnarray*}
\varphi_i(v) = \frac{v(i)}{\sum_{k \in N} v(k)} v(N) = P_i(v). 
\end{eqnarray*}
\end{proof}

Relating to \emph{equal ratio for two players}, \cite{zvcf2021scw} propose another fairness axiom in minimal cooperative settings. Their axiom, called \emph{grand worth additivity}, requires that when two-player games differ only in the grand coalition worth, the solution should behave additively across these games. In other words, the combined payoffs from the two games should coincide with the payoff in the game formed by summing their grand coalition worths. While \cite{zvcf2021scw} formulate this axiom in a variable-population framework, we adapt it to a fixed-population setting. Importantly, their characterization of the proportional division value in two-player games still applies in our setting with two non-null players, and together with \emph{projection nullified-game consistency}, this provides us with an additional characterization of the proportional division value. Following \cite{zvcf2021scw}, to formulate a fixed-population version of the axiom, we restrict attention to the class $\widehat{\mathcal{V}}^{\mathbb{Q}} := \left\{v \in \widehat{\mathcal{V}}: \text{for each} \ S \subseteq N, \ v(S) \in \mathbb{Q}\right\}$. 

\medskip
\noindent\textbf{Grand-Coalition Additivity for Two Players}: For each pair $v,w \in \widehat{\mathcal{V}}^{\mathbb{Q}}$ and each pair $i,j \in N$,
\[
\text{$v(i) = w(i)$, $v(j) = w(j)$, and each $k \neq i,j$ is a null player in $v$ and $w$} \ \Rightarrow \ \varphi(v) + \varphi(w) = \varphi(v\oplus w),
\]
where $v \oplus w \in \mathcal{V}$ coincides with \(v\) (and hence with \(w\)) on all coalitions except the grand coalition, and satisfies $(v \oplus w)(N) = v(N) + w(N)$.

\begin{theorem}\label{thm:nullifiedconsistencyadd}
A solution $\varphi: \widehat{\mathcal{V}} \to \mathbb{R}^N$ satisfies \emph{projection nullified-game consistency}, \emph{strong inessential-game property}, \emph{grand-coalition additivity for two players}, and \emph{grand-coalition continuity} if and only if $\varphi = P$.
\end{theorem}

Since the proofs of Proposition 3 and Theorem 7 in \cite{zvcf2021scw}, which characterizes the proportional division value in two-player games based on grand worth additivity, also applies in our setting, we omit the proof of Theorem \ref{thm:nullifiedconsistencyadd}.

We can also define a weaker variant of \emph{projection nullified-game consistency}, in which all players except two are nullified, and refer to it as \emph{bilateral projection nullified-game consistency}. Notably, even under this weaker condition, the two characterizations of the proportional division value still hold.

We conclude this section by providing examples to demonstrate the independence of the axioms used in Theorem \ref{thm:nullifiedconsistencyer}. We first drop \emph{projection nullified-game consistency}. Define $\varphi$ as follows: for each $v \in \widehat{\mathcal{V}}$, if there are at most $|N|-3$ null players, set $\varphi(v) = Sh(v)$;\footnote{$Sh$ denotes the Shapley value.} otherwise, set $\varphi(v) = P(v)$. Because both the Shapley value and the proportional division value satisfy \emph{efficiency}, so does $\varphi$. For each game, when at most two players are not null, the proportional division value is applied. Thus $\varphi$ satisfies \emph{equal ratio for two players}. However, because the Shapely value violates \emph{projection nullified-game consistency}, so does $\varphi$. 

We next drop \emph{efficiency}. Let $\alpha \in \mathbb{R}_{>0}$. Define the solution $\varphi$ as follows: for each $v \in \widehat{\mathcal{V}}$, set $\varphi(v) = P(v^\alpha)$. Because the proportional division value satisfies \emph{projection nullified-game consistency} and \emph{equal ratio for two players}, so does $\varphi$. Because there is $v \in \widehat{\mathcal{V}}$ such that $v(N) \neq \alpha$, $\varphi$ violates \emph{efficiency}. 

Finally, we drop \emph{equal ratio for two players}. Define $\varphi$ as follows: fix $i \in N$, and for each $v \in \widehat{\mathcal{V}}$, set $\varphi_i(v) = v(N)$, and for each $j \neq i$, $\varphi_j(v) = 0$. Since $\varphi$ always distributes $v(N)$ entirely, it satisfies \emph{efficiency}. Moreover, because player~$i$ receives the entire $v(N)$ while all other players receive $0$, independently of coalition worths, it also satisfies \emph{projection nullified-game consistency}. On the other hand, for the same reason, $\varphi$ violates \emph{equal ratio for two players}.

\section{Conclusion}\label{sec:conclusion}

We have presented an axiomatic analysis of the proportional division value in TU-games, with a focus on fixed-population consistency. We showed that the proportional division value can be characterized using homogeneity-based, composition-based, or nullified-game consistency axioms. These results demonstrate that the proportional division value is not only intuitive but also uniquely determined when fixed-population consistency is combined with mild fairness and/or efficiency requirements.

Although our focus was on the proportional division value, the fixed-population consistency axioms introduced in this paper can be applied to other classes of solutions. In our companion paper \cite{fknt2025wp}, we study linear solutions and establish characterizations based on fixed-population consistency axioms.

The proportional division value studied in this paper is defined on the domain $\mathcal{V}$, where the sum of the players' stand-alone worths is nonzero. One could extend the definition to games in $\mathcal{V}^{all}\setminus\mathcal{V}$ in several ways. For example, when $\sum_{k\in N}v(k)=0$, one could assign the grand coalition worth equally among players.\footnote{We thank an anonymous referee for suggesting this example of an extension.} Such extensions are conceptually possible, but they require specifying how payoffs should be assigned in cases where the usual proportional formula is not well defined. Moreover, extending the value to such games may raise additional continuity issues near games where $\sum_{k\in N}v(k)=0$. Such extensions may also require corresponding adjustments to the axioms studied in this paper, since the behavior of the solution at games with $\sum_{k\in N}v(k)=0$ would have to be specified explicitly. We leave the analysis of such extensions to future work.

\begin{appendix}
\section{An Incompatibility Result}\label{appendix:incompatible}
In the context of TU-games, \emph{linearity} is a standard axiom that is often imposed, and many well-known solutions, such as the Shapley value, the CIS value, and the Egalitarian Non-Separable Contribution (ENSC) value \citep{DriessenFunaki1991}, satisfy it. Interestingly, however, no \emph{linear} solution satisfies \emph{grand-coalition normalization} on the entire domain of TU-games. Note that \emph{grand-coalition normalization} is well-defined on $\mathcal{V}^{all}$. \emph{Linearity} is an algebraic axiom that requires the domain of games to be closed under addition and scalar multiplication.
The restricted domain $\mathcal{V}$ (e.g., excluding games with zero total stand-alone worth) is not closed under these operations. For this reason, it is more natural and mathematically consistent to formulate linearity on $\mathcal{V}^{all}$ rather than on $\mathcal{V}$, ensuring that sums and scalar multiples of games remain within the domain.\footnote{\textbf{Linearity}: For each pair $v,v' \in \mathcal{V}^{all}$ and each $\alpha \in \mathbb{R}$, $\varphi(v+v') = \varphi(v) + \varphi(v')$ and $\varphi(\alpha v) = \alpha \varphi(v)$.}

\begin{proposition}\label{lemma:gn&l}
\emph{Grand-coalition normalization} and \emph{linearity} are incompatible on $\mathcal{V}^{all}$.  
\end{proposition}

\begin{proof}
Suppose that there is a solution $\varphi$ satisfying \emph{grand-coalition normalization} and \emph{linearity}. By \emph{grand-coalition normalization} and \emph{linearity}, for each $v \in \mathcal{V}^{all}$ and each $i \in N$, 
\begin{eqnarray*}
\varphi_i\left(v^{v(N)-\sum_{k \in N}v(k)}\right) = \varphi_i(v) - v(i) 
&\iff& 
\varphi_i\left(v^{v(N)-\sum_{k \in N}v(k)}\right) - \varphi_i(v)=  - v(i) \\
&\iff&
\varphi_i\left(v^{v(N)-\sum_{k \in N}v(k)} - v \right) = - v(i)\\
&\iff&
\varphi_i\left(w\right) = - v(i),
\end{eqnarray*}
where 
$$
w(S)=\begin{cases}
0 \ &\text{if} \ S \neq N \\
-\sum_{k \in N}v(k)~\ &\text{if} \ S=N.
\end{cases}.
$$
Note that $w$ depends only on $-\sum_{k \in N} v(k) \in \mathbb{R}$, i.e., the total stand-alone worth. This implies that for each pair $v, v' \in \mathcal{V}$ such that $\sum_{k \in N} v({k}) = \sum_{k \in N} v'(k)$, even if there is $k \in N$ such that $v(k) \neq v'(k)$, we must have that, for each $i \in N$, $v'(i)=\varphi_i(w)=v(i)$. This is impossible. 
\end{proof}

\section{Discussion on  other bankruptcy-to-TU game constructions}
\label{sec:btoU}
A standard construction is the \emph{optimistic} mapping proposed by \cite{driessen1998greedy},\footnote{\cite{driessen1998greedy} refers to the resulting TU-game as a \emph{greedy bankruptcy game}.} 
which specifies a TU-game by assigning coalition values according to
\[
v^{(c,E)}(S)=\min\left\{\sum_{k\in S} c_k,\;E\right\}
\qquad\text{for each } S\subseteq N.
\]
This class of games is contained in $\mathcal{V}^{bank}$. 
However, analysis on this subclass substantially restricts applicability of the composition axioms.
As an illustration, consider \emph{limited composition up} with the proportional division value. 
Let $N=\{1,2\}$ and consider the reference TU-game ($v^\alpha$) induced by the bankruptcy problem $(c,E)=((2,4),5)$. Then
\[
v^{(c,E)}(1)=2,\qquad v^{(c,E)}(2)=4,\qquad v^{(c,E)}(N)=5.
\]
Hence,
\[
P_1\!\left(v^{(c,E)}\right)=\frac{2}{2+4}\cdot 5=\frac{5}{3},
\qquad
P_2\!\left(v^{(c,E)}\right)=\frac{4}{2+4}\cdot 5=\frac{10}{3}.
\]
Now suppose that, in the actual problem, the grand-coalition worth is $v(N)=5.5$.
Then the residual game $U(P(v^{(c,E)}),v)$ satisfies
\[
U(P(v^{(c,E)}),v)(1)=2-\frac{5}{3}=\frac{1}{3}, 
\qquad 
U(P(v^{(c,E)}),v)(2)=4-\frac{10}{3}=\frac{2}{3},
\]
so that
\[
U(P(v^{(c,E)}),v)(1)+U(P(v^{(c,E)}),v)(2)=1,
\]
while
\[
U(P(v^{(c,E)}),v)(N)
=
v(N)-\left(P_1(v^{(c,E)})+P_2(v^{(c,E)})\right)
=
5.5-5
=
0.5.
\]
Thus, while the residual game remains in $\mathcal{V}^{bank}$, it does not belong to the optimistic class.

Another standard construction in the literature is the \emph{pessimistic} mapping proposed by \cite{o'neill1982},\footnote{\cite{o'neill1982} refers to the resulting TU-game as a \emph{modest bankruptcy game}.} 
defined by
\[
v^{(c,E)}(S)=\max\left\{0,\,
E-\sum_{k\in N\setminus S} c_k
\right\}
\qquad\text{for each } S\subseteq N.
\]
Again, analysis focusing on this class restricts applicability of the composition axioms, particularly with \emph{limited composition up}. 
To see this, note that for the pessimistic game $v^{(c,E)}$ one has
\[
v^{(c,E)}(N)=E \ \ge\ \sum_{k\in N} v^{(c,E)}(k).
\]
Consider a solution $\varphi$ satisfying efficiency.\footnote{We provide the definition of efficiency in Section~\ref{sec:nullifiedconsistency}.} 
If $v^{(c,E)}(N)>\sum_{k\in N} v^{(c,E)}(k)$, then efficiency implies
\[
\sum_{k\in N}\varphi_k\!\left(v^{(c,E)}\right)=v^{(c,E)}(N)>\sum_{k\in N} v^{(c,E)}(k),
\]
so there exists $i\in N$ such that $\varphi_i\!\left(v^{(c,E)}\right)>v^{(c,E)}(i)$. 
By construction of the residual game,
\[
U\!\left(\varphi\!\left(v^{(c,E)}\right),\,v^{(c,E)}\right)(i)
=
v^{(c,E)}(i)-\varphi_i\!\left(v^{(c,E)}\right)
<0,
\]
and hence the residual game falls outside the pessimistic class. 
Therefore, \emph{limited composition up} is inapplicable except in the knife-edge case
\[
v^{(c,E)}(N)=\sum_{k\in N} v^{(c,E)}(k).
\]
A similar argument shows that \emph{grand-coalition normalization} is not generally applicable to this class either.

\section{Proof of Claim~\ref{claim:lcu_induction} Using \emph{Limited Composition Down}}
\label{appendix:LCD}
Analogous to the proof using \emph{limited composition down}, we argue by induction on $n$. Fix $n > 0$ and assume the induction hypothesis stated above. Let $m \in S_n \setminus S_{n-1}$, $v \in \mathcal{V}^{bank}$, and $i \in N$. 

\noindent\textbf{Case 1}: $m \leq \frac{1}{2}$. By Claim \ref{claim:lcu_101/2}, the equality holds when $m = \frac{1}{2}$. Assume now that $m < \frac{1}{2}$. By \emph{limited composition down}, $$ \varphi_i\left(v^{m\sum_{k \in N}v(k)}\right) = \varphi_i\left(D\left(\varphi\left(v^{\frac{1}{2}\sum_{k \in N}v(k)}\right),v^{m\sum_{k \in N}v(k)}\right)\right). $$ By Claim \ref{claim:lcu_101/2}, $\varphi_i\left(v^{\frac{1}{2}\sum_{k \in N}v(k)}\right) = \frac{1}{2}v(i)$. Then, \begin{eqnarray*} D\left(\varphi\left(v^{\frac{1}{2}\sum_{k \in N}v(k)}\right),v^{m\sum_{k \in N}v(k)}\right)(S) &=& \begin{cases} \sum_{k \in S} \frac{1}{2} v(k) \ &\text{if} \ S \neq N, \\ m\sum_{k \in S} v(k) \ &\text{if} \ S = N, \end{cases}\\ &=& \begin{cases} \frac{1}{2}\sum_{k \in S} v(k) \ &\text{if} \ S \neq N, \\ m\sum_{k \in S} v(k) \ &\text{if} \ S = N. \end{cases} \end{eqnarray*} In particular, for each $k \in N$, $$ D\left(\varphi\left(v^{\frac{1}{2}\sum_{k \in N}v(k)}\right),v^{m\sum_{k \in N}v(k)}\right)(k) = \frac{1}{2}v(k). $$ Then $$ D\left(\varphi\left(v^{\frac{1}{2}\sum_{k \in N}v(k)}\right),v^{m\sum_{k \in N}v(k)}\right)(N) = 2m \sum_{k \in N}\left[D\left(\varphi\left(v^{\frac{1}{2}\sum_{k \in N}v(k)}\right),v^{m\sum_{k \in N}v(k)}\right)(k)\right]. $$ Hence, $D\left(\varphi\left(v^{\frac{1}{2}\sum_{k \in N}v(k)}\right),v^{m\sum_{k \in N}v(k)}\right) \in \mathcal{V}^{bank}$ and $2m \in S_{n-1}$, so by the induction hypothesis, $$ \varphi_i\left(D\left(\varphi\left(v^{\frac{1}{2}\sum_{k \in N}v(k)}\right),v^{m\sum_{k \in N}v(k)}\right)\right) = 2m \cdot \frac{1}{2}v(i) = mv(i). $$

\noindent\textbf{Case 2}: $m > \frac{1}{2}$. By \emph{reverse grand-coalition normalization},
$$
\varphi_i\left(v^{m\sum_{k \in N}}\right) = v(i) - \varphi_i\left(v^{\left(1-m\right)\sum_{k \in N}v(k)}\right).
$$
Since $m > \frac{1}{2}$, Case 1 implies $\varphi_i\left(v^{\left(1-m\right)\sum_{k \in N}v(k)}\right) = (1-m)v(i)$, and hence
$$
\varphi_i\left(v^{m\sum_{k \in N}}\right) = v(i) - (1-m)v(i) = mv(i).
$$
\end{appendix}

\bibliographystyle{ecta}
\bibliography{references}

\begin{thebibliography}{29}
\newcommand{\enquote}[1]{``#1''}
\expandafter\ifx\csname natexlab\endcsname\relax\def\natexlab#1{#1}\fi

\bibitem[\protect\citeauthoryear{Chambers}{Chambers}{2006}]{chambers2006geb}
\textsc{Chambers, C.~P.} (2006): \enquote{Asymmetric rules for claims problems
  without homogeneity,} \emph{Games and Economic Behavior}, 54, 241--260.

\bibitem[\protect\citeauthoryear{Chambers and Moreno-Ternero}{Chambers and
  Moreno-Ternero}{2017}]{chambersmorenoternero2017scw}
\textsc{Chambers, C.~P. and J.~D. Moreno-Ternero} (2017): \enquote{Taxation and
  poverty,} \emph{Social Choice and Welfare}, 48, 153--175.

\bibitem[\protect\citeauthoryear{Dietzenbacher, Tamura, and
  Thomson}{Dietzenbacher et~al.}{2024}]{dtt2024scw}
\textsc{Dietzenbacher, B., Y.~Tamura, and W.~Thomson} (2024):
  \enquote{Partial-implementation invariance and claims problems,} \emph{Social
  Choice and Welfare}, 63, 203--229.

\bibitem[\protect\citeauthoryear{Driessen and Funaki}{Driessen and
  Funaki}{1991}]{DriessenFunaki1991}
\textsc{Driessen, T. and Y.~Funaki} (1991): \enquote{Coincidence of and
  collinearity between game theoretic solutions,} \emph{OR Spektrum}, 13,
  15--30.

\bibitem[\protect\citeauthoryear{Driessen}{Driessen}{1998}]{driessen1998greedy}
\textsc{Driessen, T.~S.} (1998): \enquote{The greedy bankruptcy game: an
  alternative game theoretic analysis of a bankruptcy problem,} \emph{Game
  theory and applications}, 4, 45--61.

\bibitem[\protect\citeauthoryear{Funaki}{Funaki}{1996}]{funaki1996wp}
\textsc{Funaki, Y.} (1996): \enquote{Dual axiomatizations of solutions of
  cooperative games,} \emph{Working Paper}.

\bibitem[\protect\citeauthoryear{Funaki and Koriyama}{Funaki and
  Koriyama}{2025}]{fk2025ijgt}
\textsc{Funaki, Y. and Y.~Koriyama} (2025): \enquote{Deriving egalitarian and
  proportional principles from individual monotonicity,} \emph{International
  Journal of Game Theory}, 54, 24.

\bibitem[\protect\citeauthoryear{Funaki, Koriyama, Nakada, and Tamura}{Funaki
  et~al.}{2026}]{fknt2025wp}
\textsc{Funaki, Y., Y.~Koriyama, S.~Nakada, and Y.~Tamura} (2026):
  \enquote{Characterizing the ELS Values with Fixed-Population Invariance
  Axioms,} \emph{Journal of Mathematical Economics}, 125.

\bibitem[\protect\citeauthoryear{Funaki and Yamato}{Funaki and
  Yamato}{2001}]{fy2001igtr}
\textsc{Funaki, Y. and T.~Yamato} (2001): \enquote{The core and consistency
  properties: a general characterisation,} \emph{International Game Theory
  Review}, 3, 175--187.

\bibitem[\protect\citeauthoryear{Kalai}{Kalai}{1977}]{kalai1977metrica}
\textsc{Kalai, E.} (1977): \enquote{Proportional solutions to bargaining
  situations: interpersonal utility comparisons,} \emph{Econometrica},
  1623--1630.

\bibitem[\protect\citeauthoryear{Kamijo and Kongo}{Kamijo and
  Kongo}{2015}]{kk2015td}
\textsc{Kamijo, Y. and T.~Kongo} (2015): \enquote{Properties based on relative
  contributions for cooperative games with transferable utilities,}
  \emph{Theory and Decision}, 78, 77--87.

\bibitem[\protect\citeauthoryear{Kaneko and Nakada}{Kaneko and
  Nakada}{2025}]{kn2025el}
\textsc{Kaneko, T. and S.~Nakada} (2025): \enquote{Nullified-game consistency
  and axiomatizations of the Core of TU-games with a fixed player set,}
  \emph{Economics Letters}, 250, 112274.

\bibitem[\protect\citeauthoryear{Khmelnitskaya and Driessen}{Khmelnitskaya and
  Driessen}{2003}]{kd2003mmor}
\textsc{Khmelnitskaya, A.~B. and T.~S. Driessen} (2003):
  \enquote{Semiproportional values for TU games,} \emph{Mathematical Methods of
  Operations Research}, 57, 495--511.

\bibitem[\protect\citeauthoryear{Moulin}{Moulin}{1987}]{moulin1987ijgt}
\textsc{Moulin, H.} (1987): \enquote{Equal or proportional division of a
  surplus, and other methods,} \emph{International Journal of Game Theory}, 16,
  161--186.

\bibitem[\protect\citeauthoryear{Moulin}{Moulin}{2000}]{moulin2000econometrica}
---\hspace{-.1pt}---\hspace{-.1pt}--- (2000): \enquote{Priority rules and other
  asymmetric rationing methods,} \emph{Econometrica}, 68, 643--684.

\bibitem[\protect\citeauthoryear{O'Neill}{O'Neill}{1982}]{o'neill1982}
\textsc{O'Neill, B.} (1982): \enquote{A problem of rights arbitration from the
  Talmud,} \emph{Mathematical social sciences}, 2, 345--371.

\bibitem[\protect\citeauthoryear{Ortmann}{Ortmann}{2000}]{ortmann2000mmor}
\textsc{Ortmann, K.~M.} (2000): \enquote{The proportional value for positive
  cooperative games,} \emph{Mathematical Methods of Operations Research}, 51,
  235--248.

\bibitem[\protect\citeauthoryear{Plott}{Plott}{1973}]{plott1973metrica}
\textsc{Plott, C.~R.} (1973): \enquote{Path independence, rationality, and
  social choice,} \emph{Econometrica}, 1075--1091.

\bibitem[\protect\citeauthoryear{Ruiz, Valenciano, and Zarzuelo}{Ruiz
  et~al.}{1996}]{rvz1996ijgt}
\textsc{Ruiz, L.~M., F.~Valenciano, and J.~M. Zarzuelo} (1996): \enquote{The
  least square prenucleolus and the least square nucleolus. Two values for TU
  games based on the excess vector,} \emph{International Journal of Game
  Theory}, 25, 113--134.

\bibitem[\protect\citeauthoryear{Shapley}{Shapley}{1953}]{Shapley1953}
\textsc{Shapley, L.~S.} (1953): \enquote{A value for n-person games,} in
  \emph{Contributions to the Theory of Games (AM-28), Volume II}, ed. by H.~W.
  Kuhn and A.~W. Tucker, Princeton, NJ: Princeton University Press, 307--317.

\bibitem[\protect\citeauthoryear{Thomson}{Thomson}{2012}]{thomson2012ep}
\textsc{Thomson, W.} (2012): \enquote{On the axiomatics of resource allocation:
  interpreting the consistency principle,} \emph{Economics \& Philosophy}, 28,
  385--421.

\bibitem[\protect\citeauthoryear{Thomson}{Thomson}{2016}]{thomson2016el}
---\hspace{-.1pt}---\hspace{-.1pt}--- (2016): \enquote{A new characterization
  of the proportional rule for claims problems,} \emph{Economics Letters}, 145,
  255--257.

\bibitem[\protect\citeauthoryear{van~den Brink}{van~den
  Brink}{2002}]{van2002ijgt}
\textsc{van~den Brink, R.} (2002): \enquote{An axiomatization of the Shapley
  value using a fairness property,} \emph{International Journal of Game
  Theory}, 30, 309--319.

\bibitem[\protect\citeauthoryear{van~den Brink, Chun, Funaki, and Park}{van~den
  Brink et~al.}{2016}]{vcfp2016td}
\textsc{van~den Brink, R., Y.~Chun, Y.~Funaki, and B.~Park} (2016):
  \enquote{Consistency, population solidarity, and egalitarian solutions for
  TU-games,} \emph{Theory and Decision}, 81, 427--447.

\bibitem[\protect\citeauthoryear{Young}{Young}{1987}]{young1987mor}
\textsc{Young, H.~P.} (1987): \enquote{On dividing an amount according to
  individual claims or liabilities,} \emph{Mathematics of Operations Research},
  12, 398--414.

\bibitem[\protect\citeauthoryear{Young}{Young}{1988}]{young1988jet}
---\hspace{-.1pt}---\hspace{-.1pt}--- (1988): \enquote{Distributive justice in
  taxation,} \emph{Journal of Economic Theory}, 44, 321--335.

\bibitem[\protect\citeauthoryear{Zou}{Zou}{2021}]{zou2021thesis}
\textsc{Zou, Z.} (2021): \enquote{Proportional values for cooperative games,}
  \emph{PhD Thesis, Vrije Universiteit Amsterdam}.

\bibitem[\protect\citeauthoryear{Zou, van~den Brink, Chun, and Funaki}{Zou
  et~al.}{2021}]{zvcf2021scw}
\textsc{Zou, Z., R.~van~den Brink, Y.~Chun, and Y.~Funaki} (2021):
  \enquote{Axiomatizations of the proportional division value,} \emph{Social
  Choice and Welfare}, 57, 35--62.

\bibitem[\protect\citeauthoryear{Zou, van~den Brink, and Funaki}{Zou
  et~al.}{2022}]{zvf2022td}
\textsc{Zou, Z., R.~van~den Brink, and Y.~Funaki} (2022): \enquote{Sharing the
  surplus and proportional values,} \emph{Theory and Decision}, 93, 185--217.

\end{thebibliography}
\end{document}